\documentclass[fleqn,usenatbib]{mnras}
\usepackage{newtxtext,newtxmath}
\usepackage[T1]{fontenc}
\usepackage{graphicx}
\usepackage{amsmath}

\renewcommand{\theequation}{\thesubsection.\arabic{equation}}

\usepackage{savesym}
\savesymbol{tablenum}
\usepackage{siunitx}
\usepackage{bm}
\usepackage{mathtools}

\usepackage{color}

\title[MRI in a Partially Ionized Gas]{Magnetorotational Instability in a Swirling Partially Ionized Gas}

\author[A. Secunda et al.]{Amy Secunda$^1$, Peter Donnel$^2$, Hantao Ji$^{1,3}$, and Jeremy Goodman$^1$ \\
$^1$Department of Astrophysical Sciences, Princeton University, Peyton Hall, Princeton, NJ 08544, USA\\
$^2$ICFP undergraduate
program, Physics department, Ecole Normale Sup\'{e}rieure, Paris, 75005,
France\\
$^3$Princeton Plasma Physics Laboratory, Princeton, NJ 08540, USA} 

\date{Accepted XXX. Received YYY; in original form ZZZ}
\pubyear{2022}

\begin{document}
\label{firstpage}
\pagerange{\pageref{firstpage}--\pageref{lastpage}}
\maketitle

\begin{abstract} 
The magnetorotational instability (MRI) has been proposed as the method of angular momentum transport that enables accretion in astrophysical discs. However, for weakly-ionized discs, such as protoplanetary discs, it remains unclear whether the combined non-ideal magnetohydrodynamic (MHD) effects of Ohmic resistivity, ambipolar diffusion, and the Hall effect make these discs MRI-stable. While much effort has been made to simulate non-ideal MHD MRI, these simulations make simplifying assumptions and are not always in agreement with each other. Furthermore, it is difficult to directly observe the MRI astrophysically because it occurs on small scales. Here, we propose the concept of a swirling gas experiment of weakly-ionized argon gas between two concentric cylinders threaded with an axial magnetic field that can be used to study non-ideal MHD MRI. For our proposed experiment, we derive the hydrodynamic equilibrium flow and a dispersion relation for MRI that includes the three non-ideal effects. We solve this dispersion relation numerically for the parameters of our proposed experiment. We find it should be possible to produce non-ideal MRI in such an experiment because of the Hall effect, which increases the MRI growth rate when the vertical magnetic field is anti-aligned with the rotation axis. As a proof of concept, we also present experimental results for a hydrodynamic flow in an unmagnetized prototype. We find that our prototype has a small, but non-negligible, $\alpha$-parameter that could serve as a baseline for comparison to our proposed magnetized experiment, which could be subject to additional turbulence from the MRI.

\end{abstract}

\section{Introduction}
Astrophysical accretion discs require a mechanism of outward angular momentum transport in order for accretion to occur. For sufficiently well-ionized accretion discs, such as an active galactic nucleus disc, the magnetorotational instability \citep[MRI,][]{balbus91} is a robust mechanism for angular momentum transport. However, for weakly-ionized accretion discs, such as protoplanetary discs, it is still heavily debated whether the MRI is sufficient to account for observed accretion rates due to non-ideal magnetohydrodynamic (MHD) effects, such as Ohmic resistivity and ambipolar diffusion, which decouple the gas disc and magnetic field stabilizing the disc \citep[e.g. ][]{gammie96,fleming00,sano02,bai11,bai13,gressel15a}. 

In protoplanetary discs, Ohmic resistivity dominates the high-density, weakly magnetized, inner midplane. Ambipolar diffusion dominates the lower-density, more strongly magnetized, outer regions and surface layers of the disc. A third non-ideal MHD effect, the Hall effect, dominates in a regime somewhere in between the other two effects at moderate densities and magnetic field strengths. Unlike the diffusive non-ideal effects, the Hall effect has been shown analytically and in simulations to moderately enhance (suppress) the MRI when the magnetic field is anti-aligned (aligned) with the axis of rotation and \citep{wardle99,bai15}.

While the MRI has been studied extensively analytically and numerically, it is often necessary to make simplifying assumptions in order to make the problem tractable. For example, numerical simulations of protoplanetary disks are often two-dimensional \cite[e.g.,][]{bai17,yang2021}, local (shearing box) approximations \citep[e.g.,][]{fleming2003,bai13,simon13,simon13b,lesur14,bai14,simon15,bai15}, of limited extent in one dimension \citep[e.g.,][]{cui2021}, and/or subject to numerical dispersion \citep[e.g.,][]{bethune17,bai17}. In addition, very few simulations of non-ideal MHD MRI include the Hall effect, and due to the computationally expensive nature of these simulations, wider parameter studies are not always feasible.

All simulations require some simplification of the physics and of the numerics and disentangling the effects of these simplifications on the results can be subtle and difficult. It is also difficult to directly observe the MRI astrophysically because it occurs on small scales. Therefore, to better understand the MRI, especially the standard version of it, or SMRI, when a magnetic field is applied along the rotation axis, several laboratory experiments have been proposed or attempted to generate the MRI that arises in a (quasi-)Keplerian flow or to generate its analogues \citep[e.g. ][see also \cite{ji23} for a recent review]{ji01,sisan04,boldyrev09,nornberg10,roach12,collins12,vasil15,ybai15,caspary18,hung19,flanagan20}. \cite{wang22a} recently successfully produced the SMRI in a Taylor-Couette cell with rotating magnetized liquid metal. This experiment was able for the first time to experimentally confirm the existence of an instability which up until this point had never been directly observed in nature, only derived theoretically. However, the conditions of these experiments are still far removed from the conditions of any astrophysical disc. 

In this paper, we propose the concept of a swirling gas experiment of partially ionized argon gas between two concentric cylinders and threaded with an axial magnetic field. Our proposed experiment will have a neutral number density and temperature that falls within the values for a minimum mass solar nebula \citep{hayashi81}, although a protoplanetary disc is primarily composed of hydrogen and helium. Another key difference is that our experimental disc will have a much higher ionization fraction of $\chi_{\rm i}\approx 10^{-3}$ as opposed to the ionization fraction of $\chi_{\rm i} \approx 10^{-13}$ in protoplanetary discs \citep{Lesur+2022}. Nonetheless, the advantage of our proposed experiment is that we should be able to probe the ambipolar-, Hall-, and Ohmic-dominated regimes in order to study how non-ideal effects suppress or enhance the MRI. The MRI has never been directly observed astrophysically or experimentally in a poorly-ionized gas. Doing so would allow for comparison with analytic predictions and numerical simulations and provide insight on what physics is most crucial to include in simulations that can be very computationally expensive.

We first derive the hydrodynamic equilibrium conditions of our experiment in Section \ref{sec:hydro}. Next, we derive a dispersion relation for non-ideal MHD MRI, describe our numerical solution, and present our parameter-dependent predictions for producing the MRI in our experiment in Section \ref{sec:mhd}. In Section \ref{sec:prototype} we describe the setup of and experimental results from a hydrodynamic prototype containing air instead of argon to validate the concept. Finally, we summarize our results in Section \ref{sec:discuss}.

\setcounter{equation}{0}
\section{Hydrodynamic equilibrium flow}
\label{sec:hydro}

\begin{figure}
    \centering
    \includegraphics[width=\columnwidth]{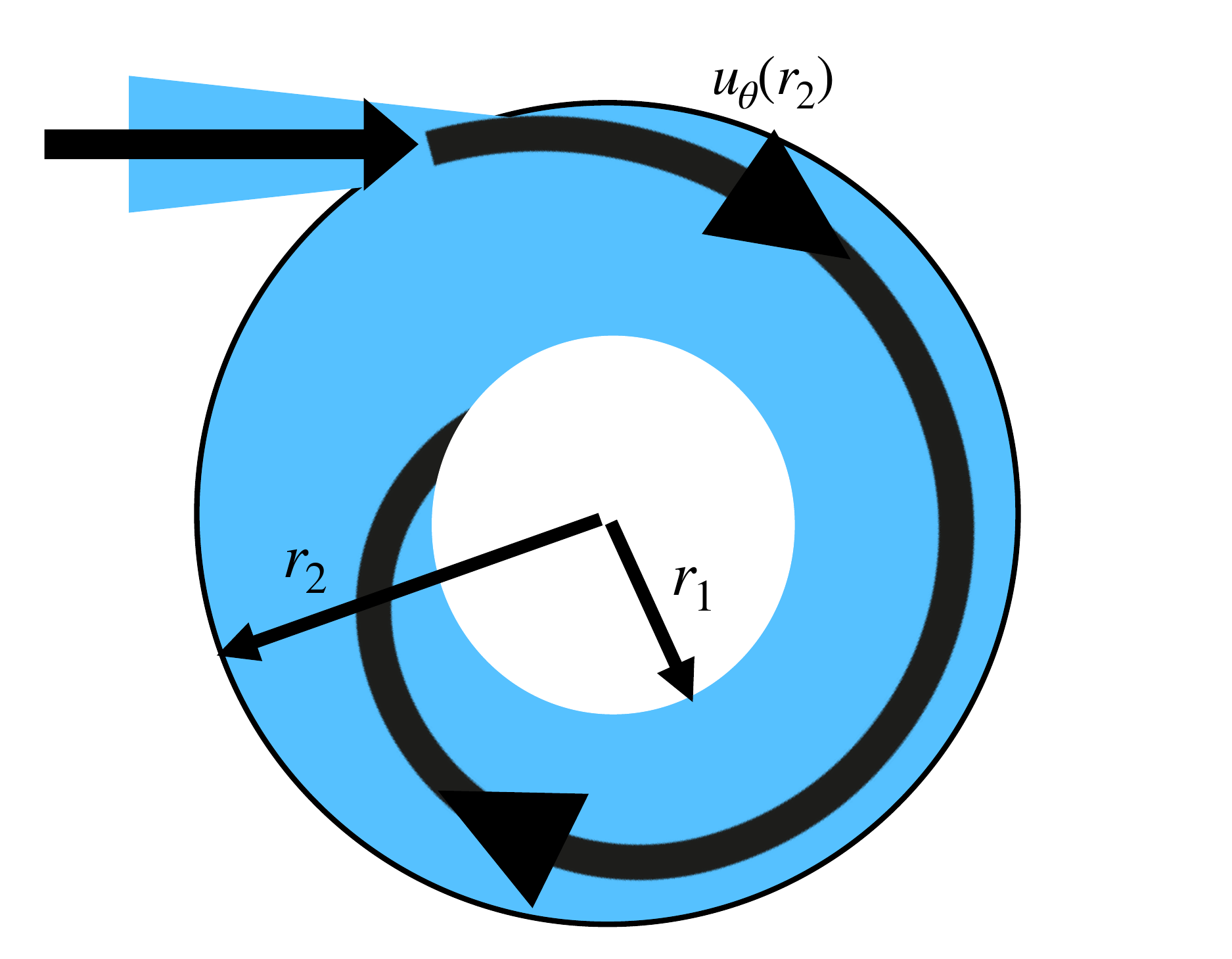}
    \caption{This cartoon of our proposed swirling gas experiment shows the top view of the swirling gas between two concentric cylinders. The gas enters the apparatus at the top of the diagram through a thin opening at an azimuthal velocity $u_{\rm \theta}(r_2)$. It spirals inward from the outer cylinder wall at $r_2$ towards the inner cylinder at $r_1$ with a radial velocity $u_{\rm r}(r)$ due to a pressure gradient imposed by a fan in the inner cylinder. The gas passes out of the apparatus through holes in the inner cylinder.}
    \label{fig:cartoon}
\end{figure}

We show a cartoon version of a top view of our proposed experiment in Figure \ref{fig:cartoon}. In the steady-state setup of our cylindrical experiment, gas is injected at the outer radius, $r_2$, with a large velocity of $u_{\rm \theta}(r_2)$ that is mostly tangential to the outer cylinder. The gas then swirls radially inwards with a radial velocity $u_{\rm r}<0$ due to a pressure gradient imposed by a fan in the inner cylinder. Finally, the gas reaches the inner cylinder with radius $r_1$ where it gets pumped out. 

Note that gas is injected into the experiment at a prescribed rate. In a steady state, the gas is also pumped out at this same rate. However, the radial structures of the swirling gas are determined by the internal dynamics of the flow. A large effective viscosity would allow rapid angular momentum transport leading to a short residual time for gas to stay in the experiment. On the other hand, a small effective viscosity would hinder angular momentum transport leading to a long gas residual time. Therefore, just as in accretion discs, an effective viscosity can be inferred from the radial profiles of the flow. Studying this effective viscosity, including that due to MRI, is the goal of our experiment.

In this section we derive the radial pressure and velocity profiles of the steady state hydrodynamic equilibrium flow in our proposed experiment. We use cylindrical coordinates ($r$, $\theta$, $z$), and assume any $\theta$ or $z$ dependencies are negligible ($\partial/\partial \theta=\partial/\partial z = 0$), which we expect to hold true far from the edges of the apparatus. We also assume the vertical component of the velocity, $u_{\rm z}$, is negligible. While the mean $u_{\rm z}$ should be roughly zero, there could be sizeable fluctuations in $u_{\rm z}$ due to turbulence, especially in the Ekman layers. However, we will assume this turbulent velocity is negligible relative to the bulk velocity. In addition we assume that the gas temperature in the device is homogeneous. In the magnetized case discussed in Section \ref{sec:mhd}, the electrons and ions will be hotter than the neutrals, but our plasma will be so poorly-ionized the electrons and ions will be unable to sufficiently heat the neutrals, which justifies our assumption of a homogeneous temperature.

For our equation of state we use, $p = \rho C^{2}$, where  $p$ is the pressure, $\rho$ is the gas density, and $C$ is the sound speed divided by the square root of the adiabatic index ($\gamma_{\rm a}=7/5$). Unless otherwise noted, we take $C$ to be constant. Finally, we assume there is an azimuthal force density $F_{\rm \theta}<0$ acting on the azimuthal velocity. In Appendix \ref{app:ftheta} we show that $F_{\rm \theta}$ represents a viscous force against the vertical boundaries, including Ekman circulation.
 
To start, we assume that the gas is compressible. In steady state, mass conservation gives,
 \begin{equation} 
 \label{masscons1}
\frac{1}{r}\frac{\partial(r\rho u_{\rm r})}{\partial r} = 0.
\end{equation}
Thus,
\begin{equation} \label{K}
 r\rho u_{\rm r} = -K,
\end{equation}
where K is a positive constant. We estimate the radial mass flux, $\rho D$, at radius $r$ as $\rho D = -2 \pi r \rho u_{\rm r} H$, where $H$ is the height of the cylinder and $D>0$ for inward flow. Using this and equation (\ref{K}) we find,
\begin{equation} 
\label{K2}
K = \frac{D \rho}{2 \pi H}.
\end{equation}

The azimuthal component of the steady-state Navier-Stokes equation is,
\begin{equation} \label{theta1}
\rho\left[ u_{\rm r}\frac{\partial u_{\rm \theta}}{\partial r} + \frac{u_{\rm r}u_{\rm \theta}}{r}\right] = F_{\rm \theta} + \mu 
\left[\frac{1}{r}\frac{\partial}{\partial r}\left(r\frac{\partial u_{\rm \theta}}{\partial r}\right) - \frac{u_{\rm \theta}}{r^{2}}\right],
\end{equation}
where $\mu$ is the dynamic viscosity. If we divide this equation by $-K/r=\rho u_{\rm r}$ we have,
\begin{equation} \label{theta2}
\frac{\partial u_{\rm \theta}}{\partial r} + \frac{u_{\rm \theta}}{r} =  \Gamma - 
\frac{\mu}{K} \left[\frac{\partial}{\partial r}\left(r\frac{\partial u_{\rm \theta}}{\partial r}\right) - \frac{u_{\rm \theta}}{r}\right],
\end{equation}
where $\Gamma = -r F_{\rm \theta}/K$. This equation has the general solution,
\begin{equation} \label{theta3}
 u_{\rm \theta} = \frac{J}{r} + \frac{r \Gamma}{2} + \frac{a}{2-\frac{K}{\mu}}r^{1-\frac{K}{\mu}},
\end{equation}
where $J$, $\Gamma$, and $a$ are all constants. We provide the derivation of this solution in Appendix \ref{app:sol}. We also show in Appendix \ref{app:sol} that in both limiting cases, $K \ll \mu$ and $\mu \ll K$, we have 
\begin{equation} \label{uthetaidk}
u_{\rm \theta} = \frac{J}{r} + \frac{r \Gamma}{2},
\end{equation}
where the definition of $\Gamma$ can be adjusted if needed.
If $K/\mu\approx2$, the final term of \eqref{theta3} should be replaced by $a r^{-1}\ln (r/r_1)$.

The radial component of the steady-state Navier-Stokes equation is,
\begin{equation} \label{radial1}
u_{\rm r}\frac{\partial u_{\rm r}}{\partial r} - \frac{u_{\rm \theta}^{2}}{r} = -\frac{1}{\rho}\frac{\partial p}{\partial r}
- \frac{\mu u_{\rm r}}{K} \left[\frac{\partial}{\partial r}\left(r\frac{\partial u_{\rm r}}{\partial r}\right) - \frac{u_{\rm r}}{r}\right].
\end{equation}
Using the radial derivative of our equation of state,
\begin{equation} \label{radial2}
\frac{1}{\rho}\frac{\partial p}{\partial r} = C^{2}\frac{\partial \ln (\rho)}{\partial r}.
\end{equation}
and the logarithmic derivative of equation (\ref{K}),
\begin{equation} \label{radial3}
\frac{\partial \ln(\rho)}{\partial r} = -\frac{1}{r} - \frac{\partial \ln(u_{\rm r})}{\partial r},
\end{equation}
in equation (\ref{radial1}) we derive,
 \begin{equation} \label{radial4}
 \begin{split}
\frac{\partial}{\partial r}\left(\frac{1}{2}u_{\rm r}^{2} - C^{2}\ln(u_{\rm r})\right) 
 + \frac{\mu u_{\rm r}}{K}
 \left[\frac{\partial}{\partial r}\left(r\frac{\partial u_{\rm r}}{\partial r}\right) - \frac{u_{\rm r}}{r}\right] \\
 = \frac{1}{r}\left(u_{\rm \theta}^{2}+C^{2}\right).
  \end{split}
 \end{equation}
From equation (\ref{K2}) we have,
 \begin{equation} \label{Kmu}
 \frac{K}{\mu} = \frac{D}{2 \pi H \nu}, 
 \end{equation}
 where the kinematic viscosity, $\nu = \mu/\rho$. Anticipating the radial flow will be larger than the viscosity for parameters of interest, we now take the limit $\mu \ll K$ in equations (\ref{theta3}) and (\ref{radial4}) and,
\begin{equation} \label{ur1}
 \frac{\partial}{\partial r}\left(\frac{1}{2}u_{\rm r}^{2} - C^{2}\ln(u_{\rm r})\right) = \frac{1}{r}\left[\left(\frac{J}{r} + \frac{r \Gamma}{2}\right)^{2}+C^{2}\right].
\end{equation}

Defining the dimensionless variables, $U \equiv u_{\rm r}/C$, $V \equiv u_{\rm \theta}/C$, $j \equiv J/r_{\rm 2}C$, $g \equiv \Gamma r_{\rm 2}/C$ and $R \equiv r/r_{\rm 2}$, where $r_{\rm 2}$ is the outer cylinder radius, equation (\ref{uthetaidk}) becomes,
\begin{equation} \label{utheta2}
V = \frac{j}{R} + \frac{g R}{2}
\end{equation}
and equation (\ref{ur1}) becomes,
\begin{equation} 
\label{ur2}
 \left(1-\frac{1}{U^{2}}\right) \frac{\partial U^{2}}{\partial R} = \frac{2}{R}\left[\left(\frac{j}{R}+\frac{g R}{2}\right)^{2}+1\right].
\end{equation}
If we define $X \equiv U^{2}$, then equation (\ref{ur2}) can be integrated over $X$ as,
\begin{equation} \label{ur3}
 X - \ln X = -\left(\frac{j}{R}\right)^2 + 2 (j g + 1)\ln R + \left(\frac{g R}{2}\right)^2 + b,
\end{equation}
where $b$ is a constant that can be estimated at $R = 1$ (i.e. $r = r_{\rm 2}$) as,
\begin{equation} \label{ur4}
b = X(1) - \ln X(1) + j^{2} - \left(\frac{g}{2}\right)^2.
\end{equation}
We can compute $X(1)$ using equations (\ref{K}) and (\ref{K2}) giving,
\begin{equation} \label{flux2}
X(1) = \left(\frac{D}{2 \pi r_{\rm 2}H C}\right)^{2},
\end{equation}
where $D$ is the flux at $r_2$.

To compare the effect of the viscosity to the effect of the geometry of the device it is useful to define the \cite{shakura73} disc viscosity parameter, $\alpha$, as
\begin{equation} \label{alpha1}
r F_{\rm \theta} = -\alpha \rho_{\rm 1} C^{2}.
\end{equation}
With this definition it is obvious that the bigger $\alpha$ is the bigger the effect of $F_{\rm \theta}$ is. If we divide this definition by $K$ and make everything dimensionless the result is,
\begin{equation} \label{alpha2}
\alpha = -\frac{g U(r_{\rm 1}) r_{\rm 1}}{r_{\rm 2}}.
\end{equation}

It is also useful to define the Reynolds number, which for a rotating flow between two concentric cylinders is defined as,
\begin{equation}
    \label{eq:re}
    Re = \frac{(r_{\rm 2}^{2}-r_{\rm 1}^{2})(\Omega_{\rm 1}-\Omega_{\rm 2})}{2 \nu}.
\end{equation}
Here $\Omega\equiv u_\theta/r$ is the angular velocity, and $\Omega_1$ and $\Omega_2$ are the angular velocities near $r_1$ and $r_2$ (but outside the boundary layers).
Equation (\ref{uthetaidk}) gives us 
\begin{equation}\label{Omega_profile}
    \Omega(r)=\Gamma/2 + J/r^2\,,
\end{equation}
so for our experiment,
\begin{equation} \label{Reynolds}
Re = \frac{J(r_{\rm 2}^{2}-r_{\rm 1}^{2})^{2}}{2\nu r_{\rm 1}^{2}r_{\rm 2}^{2} }.
\end{equation}

\begin{figure*}
    \centering
    \includegraphics[width=\textwidth]{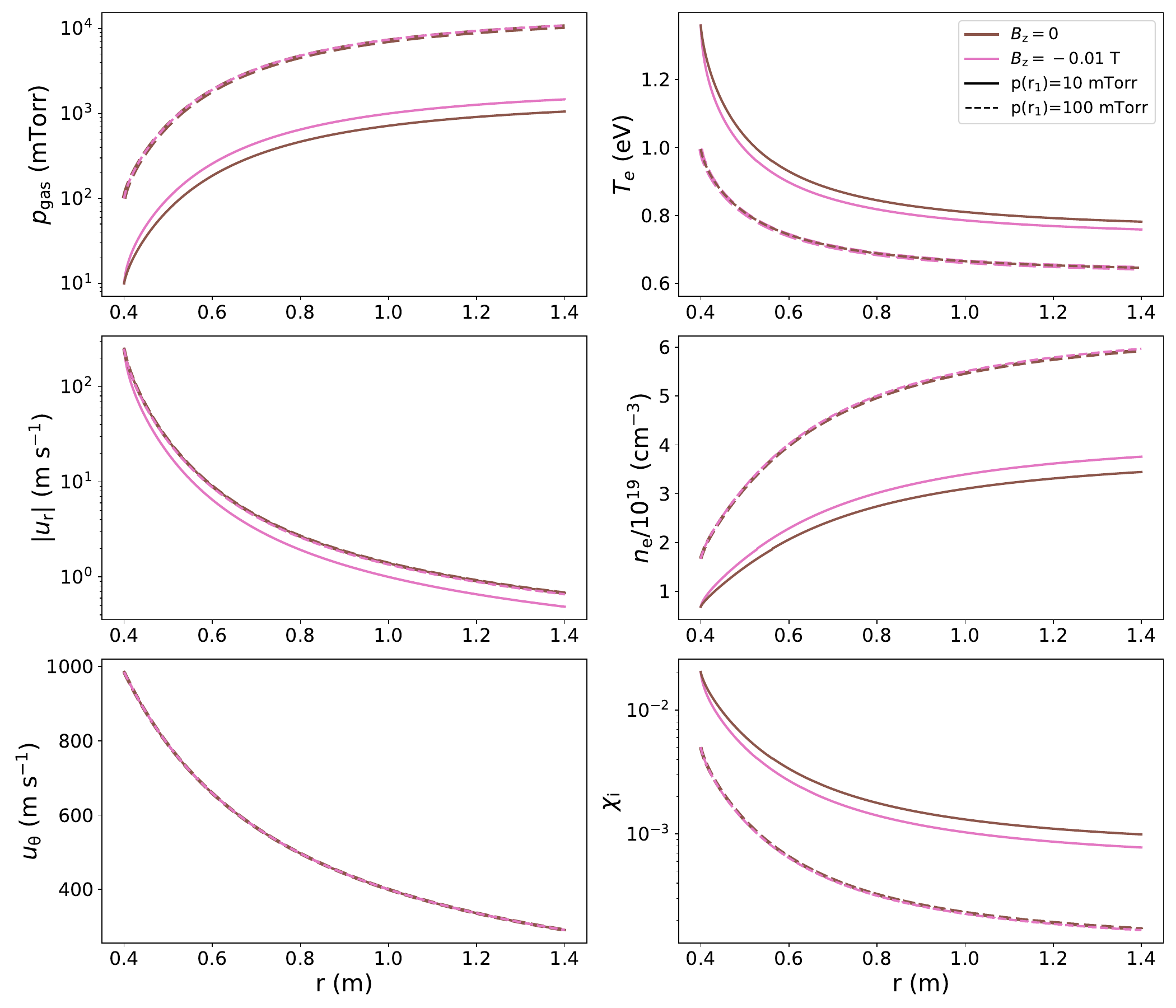}
    \caption{From top to bottom, the left panel shows the radial profiles for the gas pressure, radial velocity, and azimuthal velocity, and the right panel shows the radial profiles for the electron temperature, electron number density, and ionization fraction. The profiles for $B_{\rm z}=0$ ($B_{\rm z}=-0.01$~T) are shown in brown (pink) and the profiles for $p(r_1)=10$~mTorr ($p(r_1)=100$~mTorr) are shown as solid (dashed lines).}
    \label{fig:profs}
\end{figure*}

The left panel of Figure \ref{fig:profs} shows the radial profiles of the gas pressure, radial velocity, and azimuthal velocity for our proposed experiment, calculated using the equations in this section. The radial profiles shown in brown are for $B_{\rm z}=0$. The solid and dashed lines correspond to gas pressures, $p(r_1)=10$~mTorr and $p(r_1)=100$~mTorr, respectively. We set $\Gamma=-0.01$~s$^{-1}$, $u_{\rm r}(r_1)=-250$~m~s$^{-1}$, and $u_{\rm \theta}(r=1~{\rm m})=400$~m~s$^{-1}$. First we determine $u_{\rm \theta}(r)$ using equation (\ref{uthetaidk}). Next, we use equation (\ref{ur3}) to determine $u_{\rm r}(r)$. Because the gas temperature is constant in our calculations, once we have calculated $u_{\rm r}(r)$ we determine the gas pressure from the gas density using equation (\ref{K}), where we take $K=-u_{\rm r}(r_1)r_1p(r_1)/(k_{\rm B} T_{\rm n})$, where $k_{\rm B}$ is Boltzmann's constant and $T_{\rm n}$ is the gas temperature. 

For the chosen parameters, we calculate that the gas pressure increases by roughly two orders of magnitude from the inner to outer radius. Therefore the gas density is not constant and the assumption of incompressibility in Section \ref{sec:disp} has to be taken as a first approximation. The magnitude of the radial velocity decreases by two orders of magnitude from the inner to outer radius, because it is inversely proportional to density (see Equation (\ref{K})). For the same reason, the magnitude of the radial velocity decreases for higher $p(r_1)$. The azimuthal velocity does not depend on the pressure and consistently goes from 1000~m~s$^{-1}$ at the inner radius to 250~m~s$^{-1}$ at the outer radius.

\section{Non-ideal magnetorotational instability}
\setcounter{equation}{0}
\label{sec:mhd}
In the previous section we outlined the hydrodynamic equilibrium flow for our proposed swirling gas experiment. In this section, we add a magnetic field to this hydrodynamic flow that may or may not act to destabilize the flow. We derive a dispersion relation for MRI in the presence of Ohmic, ambipolar, and Hall effects.
Then we describe our procedures for evaluating the physical parameters of the dispersion relation numerically and solving for MRI growth rates. Using these numerical solutions, we make predictions for MRI for a wide range of parameters that are possible for a future swirling-argon-gas experiment.  

\setcounter{equation}{0}
\subsection{Deriving the Dispersion Relation}
\label{sec:disp}
The equations of non-ideal MHD are,
\begin{enumerate}
\item Continuity equation:
\begin{equation} \label{consmass3}
 \frac{\partial \rho}{\partial t} + \overrightarrow{\nabla} \bmath{\cdot} (\rho \bmath{u}) = 0
\end{equation}
\item Gauss' law for magnetism:
\begin{equation} \label{mmagmono}
 \overrightarrow{\nabla} \bmath{\cdot} \bmath{B} = 0
\end{equation}
\item Momentum equation:
\[ \rho \left(\frac{\partial \bmath{u}}{\partial t} + \left(\bmath{u} \bmath{\cdot} \overrightarrow{\nabla}\right) \bmath{u}\right) \]
\begin{equation} \label{mnavstokes}
\hspace{13mm} = - \overrightarrow{\nabla}\left(p + \frac{B^{2}}{2 \mu_{\rm 0}}\right) + \frac{\left(\bmath{B} \bmath{\cdot} \overrightarrow{\nabla} \right) \bmath{B}}{\mu_{\rm 0}} - \mu \left( \overrightarrow{\nabla} \times \left( \overrightarrow{\nabla} \times \bmath{u} \right) \right)
\end{equation}
\item The induction equation:
\[\frac{\partial \bmath{B}}{\partial t} = \overrightarrow{\nabla} \bmath{\times} \left[\bmath{u} \bmath{\times B} - \frac{\left(\overrightarrow{\nabla} \bmath{\times B}\right) \bmath{\times B}}{\mu_{\rm 0} e n_{\rm e}}\right.\]
\begin{equation} \label{dbdt}
\hspace{32mm}\left. +
\frac{\left(\bmath{j} \bmath{\times B}\right) \bmath{\times B}}{\gamma_{\rm in} \rho_{\rm i}\rho}\right]
 + \eta_{\rm o} \overrightarrow{\nabla}^2 \bmath{B},
\end{equation}
\end{enumerate}
where $\bmath{B}$ is the magnetic field, $\bmath{u}$ is the velocity vector, $\bmath{j}$ is the current, $\mu_0$ is the vacuum permeability, $e$ is the charge of an electron, $n_{\rm e}$ is the electron number density, $\gamma_{\rm in}$ is the ion-neutral collision rate, and $\eta_{\rm o}$ is the Ohmic diffusivity. In the induction equation, the first term in brackets corresponds to ideal MHD, the second term is the Hall term, which accounts for the ions having greater inertia than the electrons, and the third term is the ambipolar diffusion term, which accounts for the ion-neutral drift. The final term in the induction equation is the resistive term.

Next we introduce axisymmetric velocity, $v_{\rm r}, v_{\rm \theta}, v_{\rm z}$, magnetic field, $b_{\rm r}, b_{\rm \theta}, b_{\rm z}$, and pressure, $p_1$, perturbations of the form, $X \equiv x e^{\gamma t - ik_{\rm r}r - ik_{\rm z}z}$. We assume that the background flow is steady and incompressible, which means equation \ref{consmass3} becomes $\overrightarrow{\nabla}\bmath{\cdot u}=0$. We define
\[\bmath{u} = 
\left(
\begin{array}{c}
\frac{-K}{\rho r} + v_{\rm r} \\
\frac{\Gamma r}{2} + \frac{J}{r} + v_{\rm \theta} \\
v_{\rm z}
\end{array}
\right),
\; \; \;
\bmath{B} = 
\left(
\begin{array}{c}
b_{\rm r} \\
b_{\rm \theta} \\
B_{\rm z} + b_{\rm z}
\end{array}
\right),
\;\;\;
p = p_{\rm 0} + p_{\rm 1}.\]
Here the background flow has the radial velocity profile given by equation (\ref{K}), the azimuthal velocity profile given equation (\ref{uthetaidk}), and pressure profile $p_0(r)$. The background magnetic field is purely vertical and uniform, so that there is no background current.

If we then assume $k_{\rm r},k_{\rm z}\gg 1/r$ and linearize the MHD equations, we have 
 \begin{equation} \label{mhd1}
 k_{\rm r}v_{\rm r} + k_{\rm z}v_{\rm z} = 0,
 \end{equation}
 \begin{equation} \label{mhd2}
 k_{\rm r}b_{\rm r} + k_{\rm z}{b}_{\rm z} = 0, 
 \end{equation}
 \begin{equation} \label{mhd3}
 \left(\gamma + \eta_{\rm o} k^{2} + \eta_{\rm A} k^{2}\right)b_{\rm r} = -i k_{\rm z} B_{\rm z} v_{\rm r} - 
k_{\rm z}^{2} \eta_{\rm H} b_{\rm \theta},
 \end{equation}
 \begin{equation} \label{mhd4} 
 \left(\gamma + \eta_{\rm o} k^{2} + \eta_{\rm A} k_{\rm z}^{2}\right)b_{\rm \theta} = -i k_{\rm z} B_{\rm z} v_{\rm \theta} +
 k^{2} \eta_{\rm H} b_{\rm r} + \frac{\partial \Omega}{\partial \ln r}b_{\rm r},
 \end{equation}
 \begin{equation} \label{mhd5}  
 \left(\gamma + \nu k^{2}\right)v_{\rm r} = 2 \Omega v_{\rm \theta} -\frac{ik_{\rm z}B_{\rm z}b_{\rm r}}{\mu_{\rm 0} \rho} + \frac{i k_{\rm r}p_{\rm 1}}{\rho}
 + \frac{i k_{\rm r} B_{\rm z}{b}_{\rm z}}{\mu_{\rm 0} \rho},
 \end{equation}
 \begin{equation} \label{mhd6}
 \left(\gamma + \nu k^{2}\right)v_{\rm \theta} = - \Gamma v_{\rm r} - \frac{i k_{\rm z}B_{\rm z}b_{\rm \theta}}{\mu_{\rm 0} \rho},
 \end{equation}
 \begin{equation} \label{mhd7}
 \left(\gamma + \nu k^{2}\right)v_{\rm z} = \frac{i k_{\rm z}p_{\rm 1}}{\rho},
 \end{equation}
where ambipolar diffusivity, $\eta_{\rm A} \equiv B_{\rm z}^2/({\mu_{\rm 0} \gamma \rho_{\rm i} \rho})$, and the Hall term, $\eta_{\rm H} \equiv B_{\rm z}/(e \mu_0 n_{\rm e})$. From these equations we can derive the fourth order dispersion relation,
\[\gamma^{4} + \Big[2(\nu + \eta_{\rm o})k^{2} + \eta_{\rm A}\left(k^{2}+k_{\rm z}^{2}\right)\Big]
 \gamma^{3}\]
 \[+ \bigg\{ k^{2}\Big[\nu^{2}k^{2} + 2 \nu \left(2 \eta_{\rm o}k^{2} +\eta_{\rm A}(k^{2}+k_{\rm z}^{2})\right) + \left(\eta_{\rm o}+ \eta_{\rm A}\right)\left(\eta_{\rm o}k^{2} + \eta_{\rm A}k_{\rm z}^{2}\right)\Big]\]
\[\hspace{5mm} + 2\Omega \Gamma\frac{k^2_{\rm z}}{k^2} 
+ \eta_{\rm H}k^2_{\rm z} \Big[(\Gamma - 2 \Omega)
 + k^{2}\eta_{\rm H}\Big] + 2 k_{\rm z}^{2}V_{\rm A}^{2}\bigg\} \gamma^{2}\]
 \[+\bigg\{ \nu k^{4}\Big[\nu\left(2 \eta_{\rm o}k^{2}+\eta_{\rm A}\left(k^{2}+k_{\rm z}^{2}\right)\right)+ 2\left(\eta_{\rm o}+ \eta_{\rm A}\right)
 \left(\eta_{\rm o}k^{2} + \eta_{\rm A}k_{\rm z}^{2}\right)\Big] \] 
 \[\hspace{5mm}+ 2\Omega \Gamma\frac{k^2_{\rm z}}{k^2} \left(2 \eta_{\rm o}k^{2}+\eta_{\rm A}\left(k^{2}+k_{\rm z}^{2}\right)\right) \]
 \[\hspace{10mm}+ 2 \nu \eta_{\rm H}k^{2}k^2_{\rm z} \Big[(\Gamma - 2 \Omega)+
 \eta_{\rm H}k^{2}\Big] \]
 \[\hspace{15mm}+ k_{\rm z}^{2}V_{\rm A}^{2}\Big[2(\nu + \eta_{\rm o})k^{2} + \eta_{\rm A}\left(k^{2}+k_{\rm z}^{2}\right)\Big]\bigg\} \gamma\]
\[+ \left(k^{6} \nu^{2}+ 2 \Omega \Gamma k_{\rm z}^{2}\right)(\eta_{\rm o}+ \eta_{\rm A})\left(\eta_{\rm o}k^{2} + \eta_{\rm A}k_{\rm z}^{2}\right) \]
\[\hspace{5mm}+ \nu^{2} \eta_{\rm H}k^{4}k_{\rm z}^2 \Big[(\Gamma - 2 \Omega) + \eta_{\rm H}k^{2}\Big]\]
\[\hspace{10mm}+ \nu k^{2}k_{\rm z}^2V_{\rm A}^{2}  \Big[2 \eta_{\rm o}k^{2} +\eta_{\rm A}(k^{2}+k_{\rm z}^{2})\Big] \]
\begin{equation} \label{eq:monstergamma}
+ \frac{k_{\rm z}^{4}}{k^2}\left(2 \Omega \Gamma\eta_{\rm H} +
2\Omega V_{\rm A}^{2}\right)\Big[\frac{V_{\rm A}^2 k^{2}}{2\Omega} +  (\Gamma - 2 \Omega)+\eta_{\rm H}k^{2}\Big] = 0,
\end{equation}
where $V_{\rm A}=B_{\rm z}/\sqrt{\mu_0 \rho}$ is the Alfvén velocity, and $k^2\equiv k_{\rm r}^2+k_\theta^2$.
For a more general angular-velocity profile, $\Omega(r)$, the combinations $\Gamma$ and $\Gamma-2\Omega$ would be replaced by $r^{-1}d(r^2\Omega)/dr$ and $rd\Omega/dr$, respectively.
Notice that the latter always occurs added to $\eta_{\rm H} k^2$, hinting at the importance of the Hall effect in promoting or inhibiting instability, depending on the sign of $\eta_{\rm H}$. \cite{Kunz+Balbus2004} give a dispersion relation equivalent to equation~\eqref{eq:monstergamma}, except for the omission of the viscous terms, but allowing for an azimuthal as well as vertical component of the background field.

\setcounter{equation}{0}
\subsection{Numerical Solution}
\label{sec:numerical}

\begin{figure*}
\begin{center} 
\includegraphics[width=0.48\textwidth]{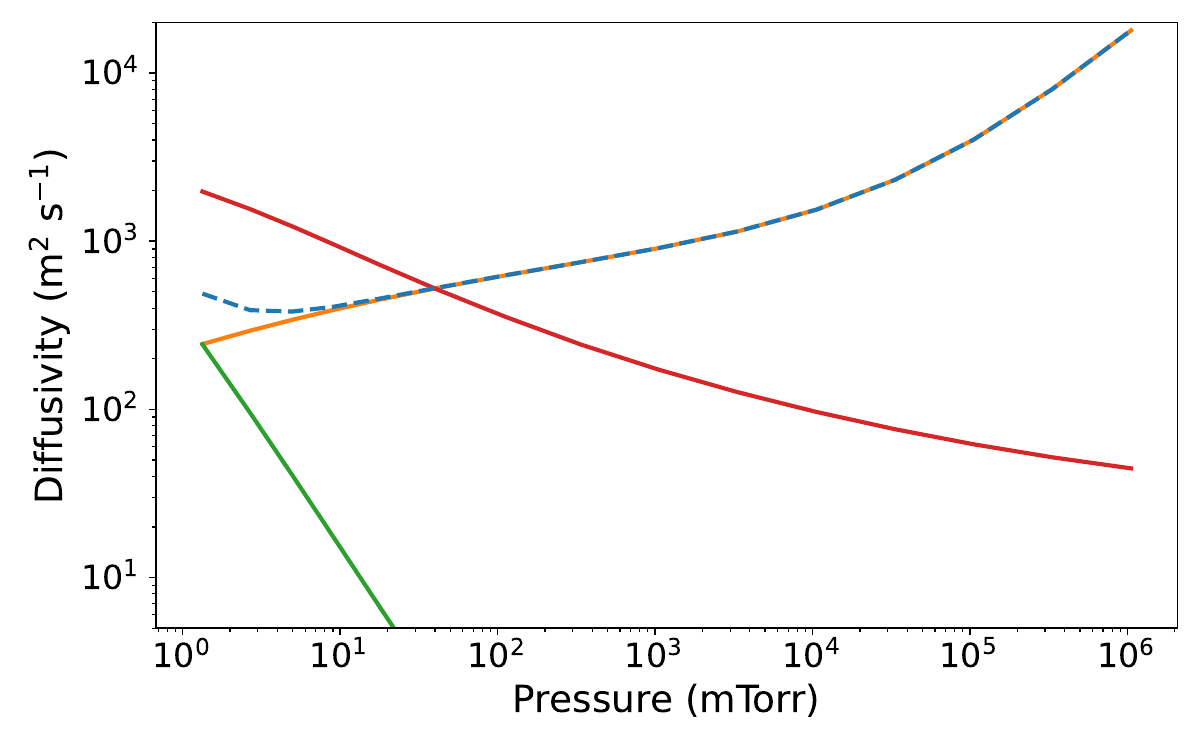}
\includegraphics[width=0.48\textwidth]{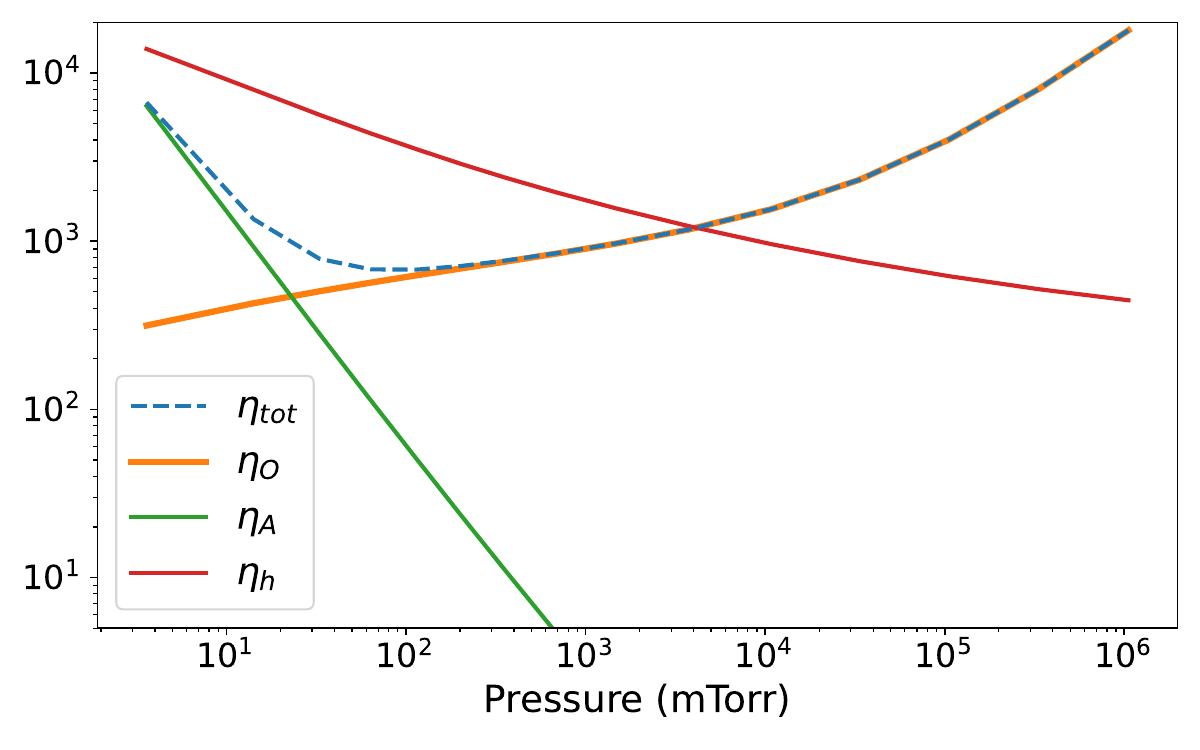}
\caption{The Ohmic ($\eta_{\rm o}$) and ambipolar ($\eta_{\rm A}$) diffusivities, as well as the Hall term ($\eta_{\rm H}$) and total diffusivity ($\eta_{\rm tot}$), where $\eta_{\rm tot} = \eta_{\rm o}+\eta_{\rm A}$, as a function of $p(r_2)$ for our fiducial setup. In the left panel $B_{\rm z}=0.001$~T and in the right panel $B_{\rm z}=0.01$~T.} 
\label{fig:diffusive}
\end{center}
\end{figure*}

\begin{figure}
\begin{center} 
\includegraphics[width=1.\linewidth]{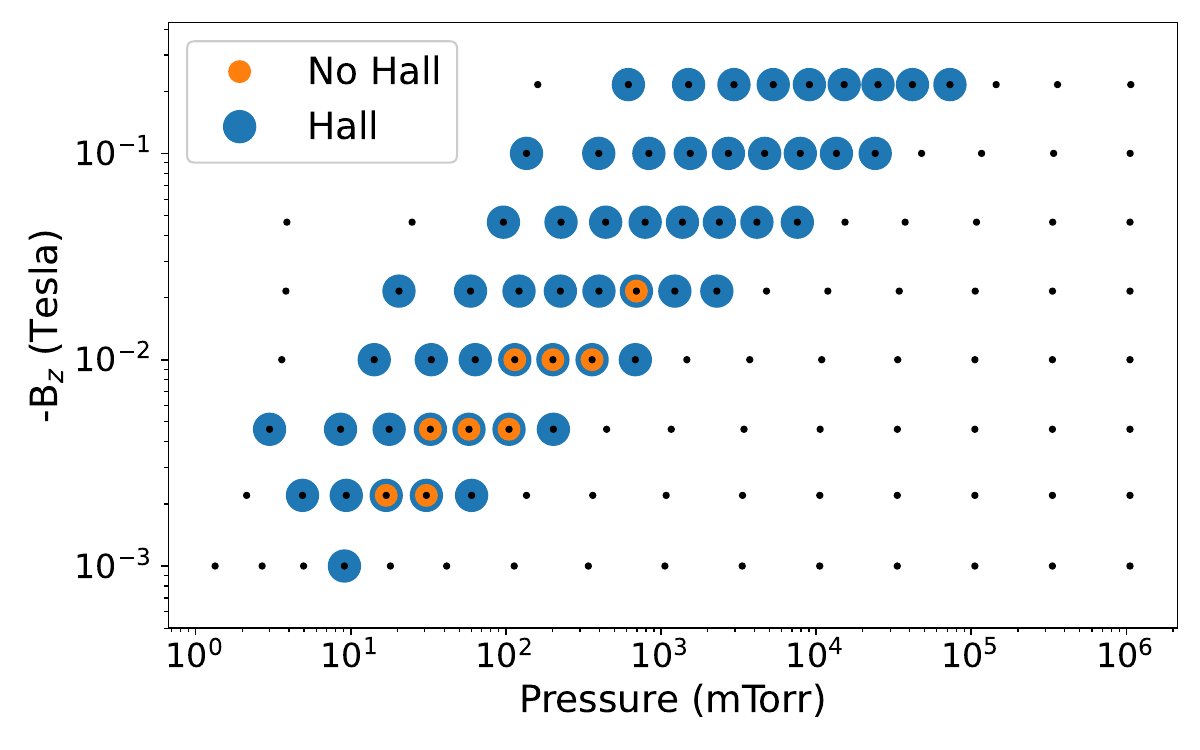}
\caption{Blue dots show pressures and magnetic field strengths for which the solution to Equation (\ref{eq:monstergamma}) is positive, i.e. there is MRI growth with the Hall effect. Orange dots show where there is MRI growth without the Hall effect (positive solutions to Equation (\ref{eq:monstergamma}) with $\eta_{\rm H}=0$). Black dots show all magnetic field strengths and pressures used in our calculation regardless of whether MRI growth is found.}
\label{fig:g_ideal}
\end{center}
\end{figure}

\begin{figure*}
    \centering
    \includegraphics[width=\textwidth]{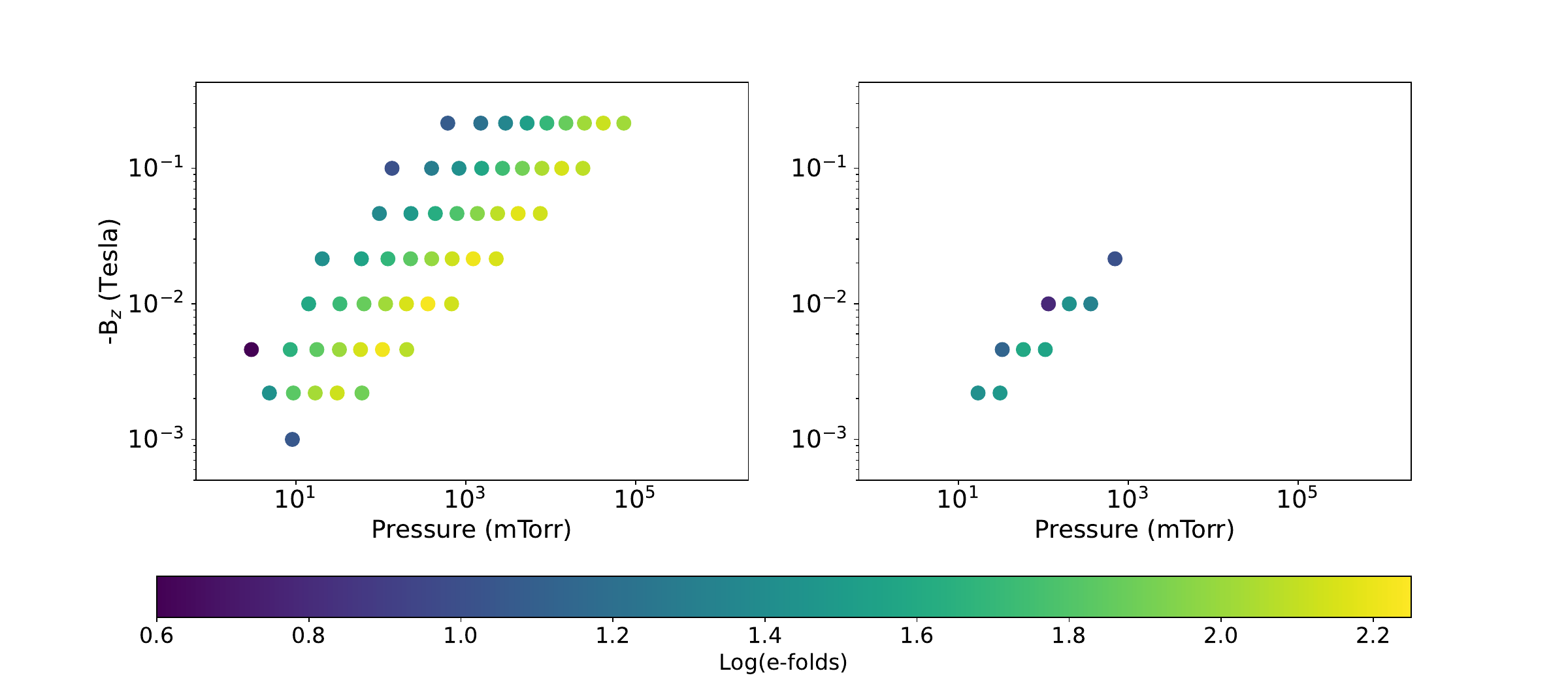}
    \caption{The logarithm of the number of e-folding times for MRI growth for the residual time of the gas in the apparatus with the Hall effect (left panel) and without the Hall effect (right panel) for a range of pressures at $r_2$ and vertical magnetic field strengths.}
    \label{fig:grate}
\end{figure*}

\begin{figure*}
    \centering
    \includegraphics[width=\textwidth]{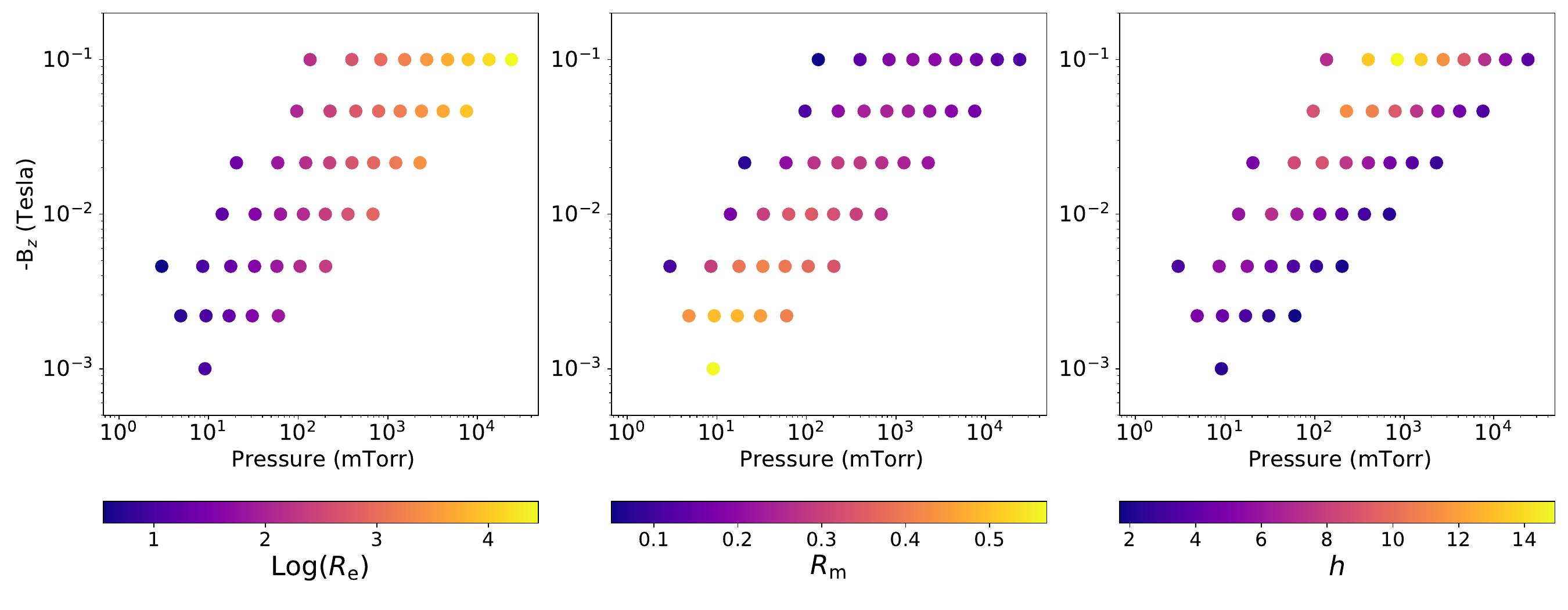}
    \caption{The logarithm of the Reynolds number (left panel), the magnetic Reynolds number (center panel), and the absolute value of the Hall parameter (right panel) as a function of pressure and magnetic field strength for parameters with MRI growth for calculations with the Hall effect.}
    \label{fig:visc}
\end{figure*}

\begin{figure}
    \centering
    \includegraphics[width=0.48\textwidth]{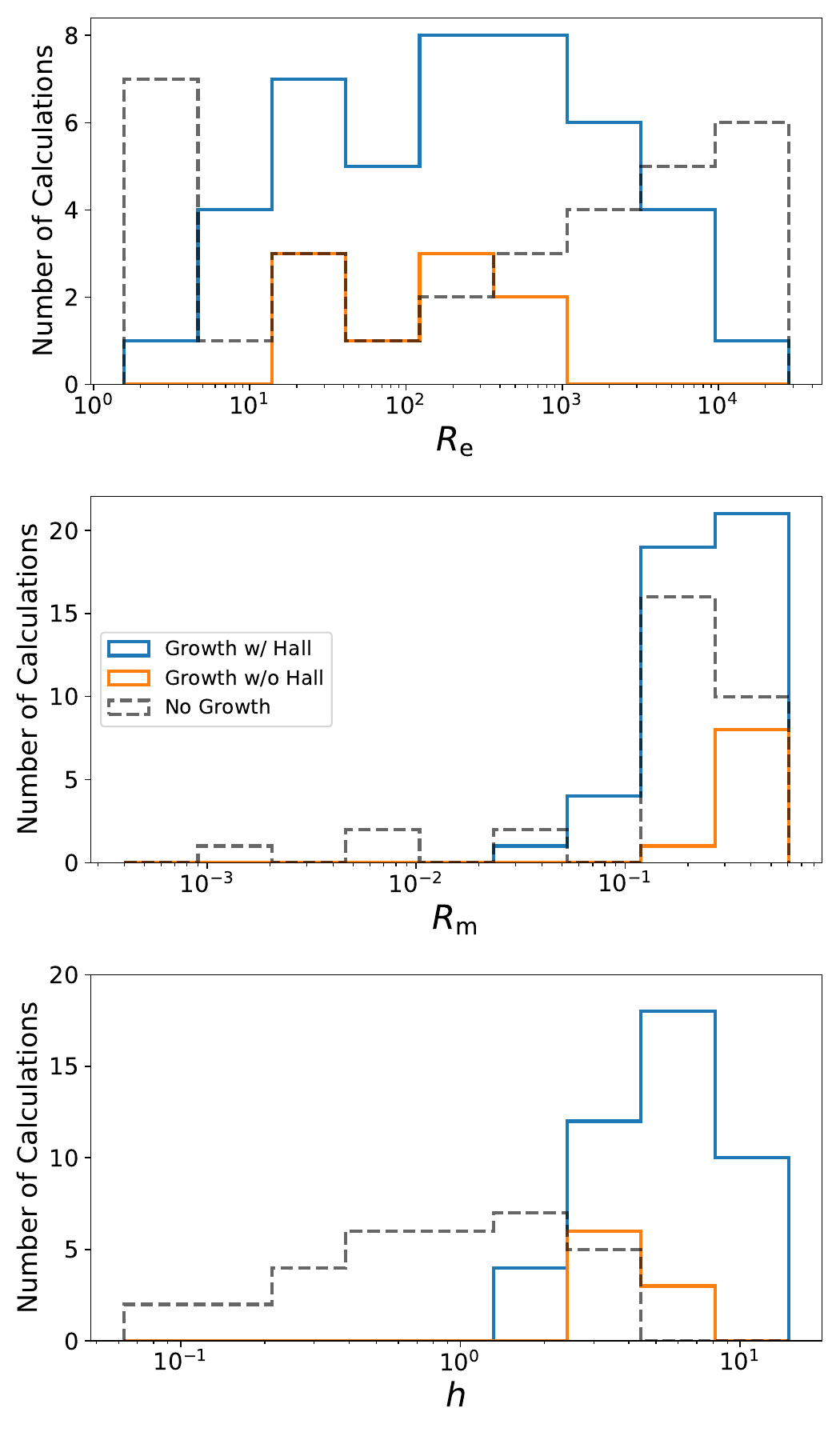}
    \caption{The distribution of Reynolds numbers (top panel), magnetic Reynolds numbers (center panel), and Hall parameters (bottom panel), for which there is MRI growth for calculations with the Hall effect (in blue), MRI growth for calculations without the Hall effect (in orange), and no MRI growth (dashed black lines in black).}
    \label{fig:visc_hist}
\end{figure}

\begin{figure}
    \centering
    \includegraphics[width=1.\linewidth]{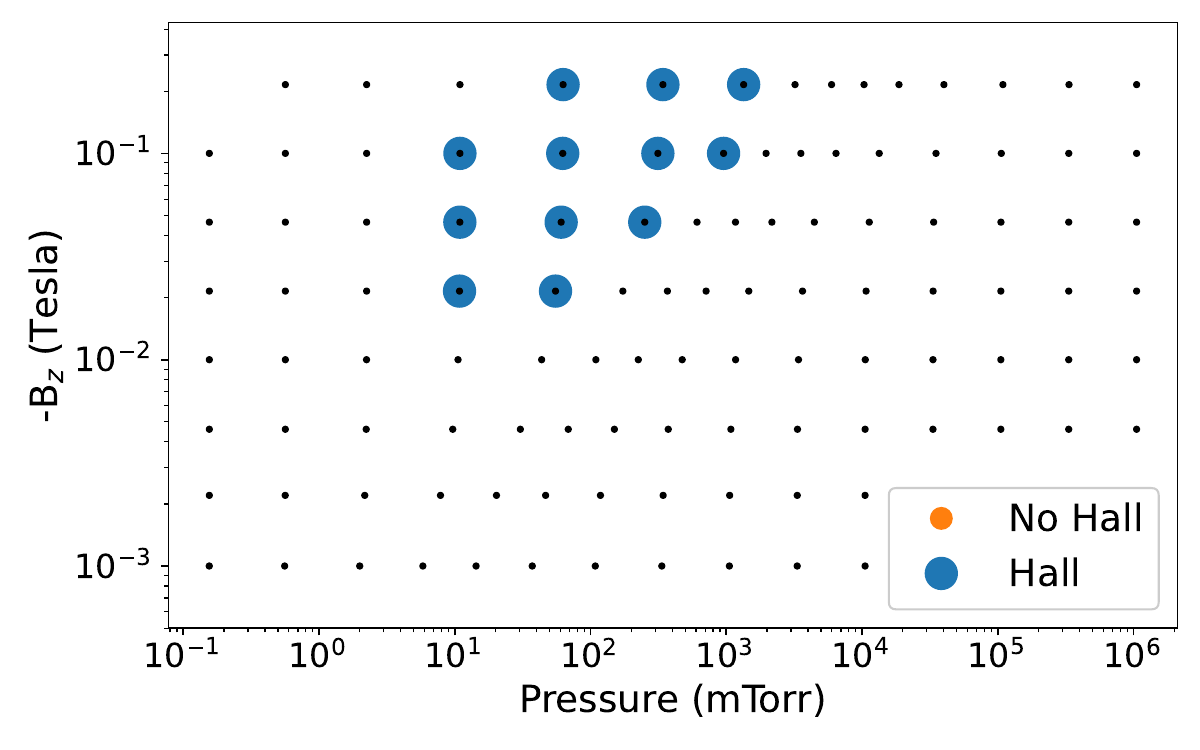}
    \includegraphics[width=1.\linewidth]{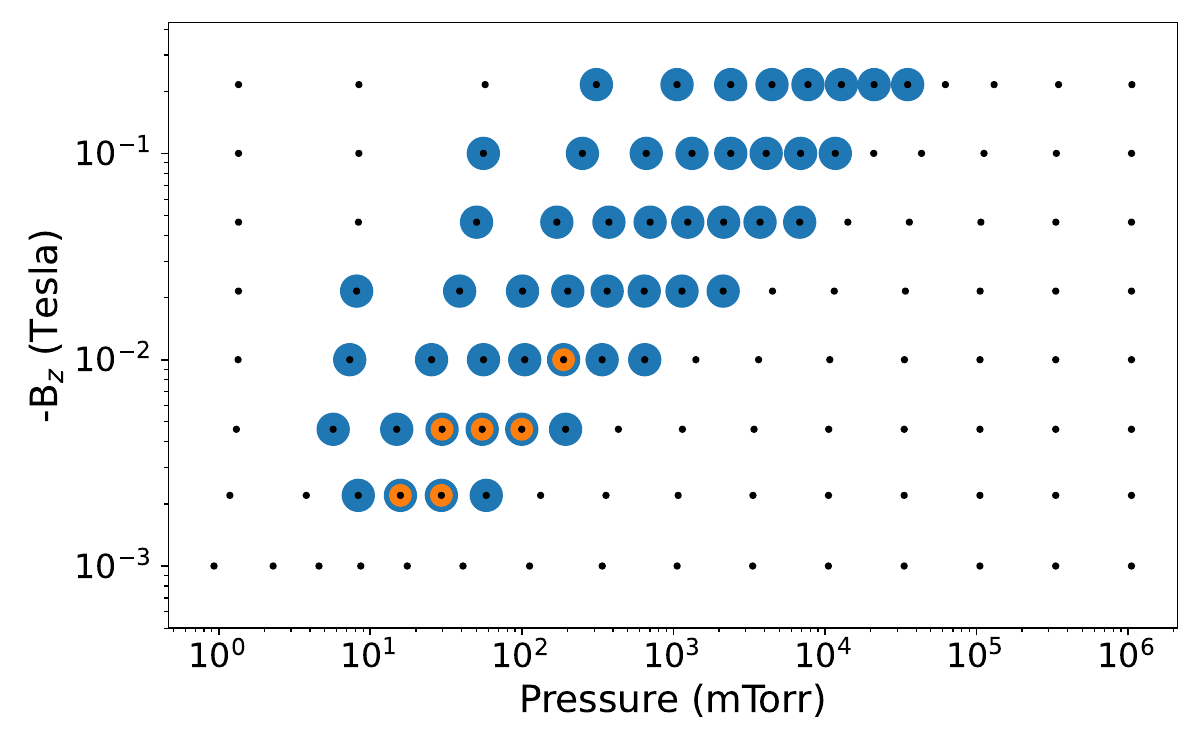}
    \caption{Same as Figure \ref{fig:g_ideal} but for two different powers: 10~kW (top panel) and 100~kW (bottom panel).}
    \label{fig:g_powers}
\end{figure}

We solve for the real roots of this dispersion relation numerically. A positive real root corresponds to MRI growth. To solve this dispersion relation we need to determine the Alvfén velocity, viscosity, Ohmic and ambipolar diffusivities, and Hall term for our proposed ionized swirling argon experiment. To find these parameters we first determine the temperature and density profiles of the experiment. 

We set the ion and gas temperatures to a constant $T_{\rm i} = T_{\rm n} = 500$~K. As in Section \ref{sec:hydro}, we calculate the radial density profile using Equation (\ref{K}), where we take $K=-u_{\rm r}(r_1)r_1\rho(r_1)$. However, when $B\neq0$, to calculate $u_{\rm r}(r)$ we need to add $j_{\theta} \times B_{\rm z}$, where $j_{\theta}=(v_{\rm r} \times B_{\rm z})/(\eta_{\rm o} + \eta_{\rm a})$ is the azimuthal component of the current, to the radial force balance equation, Equation (\ref{radial1}). Carrying this term through to Equation (\ref{ur2}), we get,
\begin{equation} 
\label{ur2B}
 \left(1-\frac{1}{U^{2}}\right) \frac{\partial U^{2}}{\partial R} = \frac{2}{R}\left[\left(\frac{j}{R}+\frac{g R}{2}\right)^{2}+1\right] + \frac{2B_{\rm z}^2 r_2^2}{(\eta_{\rm o} + \eta_{\rm a}) K}V^2 R,
\end{equation}
which we then integrate for the initial radial velocity profile. 

The electron density and temperature depend on the geometry of the device and the input power, $P$. We use the model from \cite{lieberman_2005}, outlined in Appendix \ref{app:nete}, to determine the electron temperature and density profile of our experiment. 

We show the radial profiles of the gas pressure, radial velocity, azimuthal velocity, electron temperature, electron number density, and ionization fraction in Figure \ref{fig:profs} calculated for $p(r_1)=10$~mTorr ($p(r_1)=100$~mTorr) and $B_{\rm z}=0$ and $B_{\rm z}=-0.01$~T as brown and pink solid (dashed) lines, respectively. Compared to when $B_{\rm z}=0$, for $B_{\rm z}=-0.01$~T there is an increase in both the pressure and radial velocity gradients. This increase is due to the last term in Equation (\ref{ur2B}) which is zero for $B_{\rm z}=0$. When this term is non-zero, both the pressure and radial velocity gradients must increase to compensate. When the pressure gradient increases, the neutral number density increases, which in turn increases the electron number density. However, without increasing the power, for a higher density the ionization fraction will decrease slightly, as will the electron temperature.

As the initial pressure increases the relative impact of the added $j_{\theta} \times B_{\rm z}$ term decreases. For $p(r_1)=100$~mTorr (dashed lines in Figure \ref{fig:profs}) the various radial profiles for the two magnetic field strengths are nearly identical. Increasing the gas pressure increases the electron number density, while decreasing the electron temperature and ionization fraction, because the power is kept constant. For $p(r_1)=100$~mTorr the radial velocity is similar to its value for $p(r_1)=10$~mTorr and $B_{\rm z}=0$. This similarity is because only increasing the gradient of the pressure has an affect on the radial velocity, and at higher pressures, the magnetic field is not significant enough to affect the radial velocity. Over the range of parameters we use in our calculations, the electron temperature tends to fall around $T_{\rm e}=1$~eV and the electron number density is around $n_{\rm e}=10^{19}$~m$^{-3}$. 

As a simplification, we use the radially averaged values of these temperature and density profiles to calculate $v_{\rm A}$, $\nu$, $\eta_{\rm o}$, $\eta_{\rm a}$, and $\eta_{\rm H}$. We assume the viscosity is dominated by neutral viscosity,
\begin{equation}
    \label{eq:visc}
    \nu_{\rm nn} = \lambda_{\rm nn} \gamma_{\rm nn}=\frac{v_{\rm n}^2}{\gamma_{\rm nn}}=\frac{1}{n_{\rm n}\sigma_{\rm n}} \sqrt{\frac{k_{\rm B} T_{\rm n}}{m_{\rm n}}},
\end{equation}
where $\lambda_{\rm nn}=v_{\rm n}/\gamma_{\rm nn}$ is the neutral-neutral mean free path, $\gamma_{\rm nn}=n_{\rm n}\sigma_{\rm n}\sqrt{4k_{\rm B}T_{\rm n}/m_{\rm n}}$ is the neutral-neutral collision rate, $m_{\rm n}$ is the mass of neutral argon, $n_{\rm n}$ is the neutral number density, $\sigma_{\rm n} \approx  10^{-19}$~m$^2$ is the neutral collision cross-section, and $v_{\rm n} = \sqrt{2k_{\rm B} T_{\rm n}/m_{\rm n}}$ is the neutral velocity.

We calculate Ohmic resistivity as,
\begin{equation}
    \eta_{\rm o}=\frac{m_{\rm e} n_{\rm n}}{e^2n_{\rm e}}\langle \sigma v \rangle_{\rm en}, 
\end{equation}
where $m_{\rm e}$ is the electron mass. $\langle \sigma v \rangle_{\rm en}$ is the collision rate between electrons and neutrals which depends on $\sigma_{\rm en} \approx 3.5 \times 10^{-20}$~m$^2$, the electron-neutral collision cross-section at $T_{\rm e} \approx 4$~eV \citep{Pitchford_2013}, and the electron velocity, $v_{\rm e}=\sqrt{2k_{\rm B}T_{\rm e}/m_{\rm e}}$.

\begin{table}
    \centering
    \begin{tabular}{c|c}
        Parameter & Range \\
        \hline
        $p(r_1)$ & $10^{-3}$ -- $10^4$~mTorr  \\
        $|B_z|$ & $10^{-3}$ -- 0.2~T\\
        $P$ & 1 -- 200~kW\\
        $H/(r_2 - r_1)$ & 0.5 -- 2\\
    \end{tabular}
    \caption{The parameter ranges over which we numerically solve equation (\ref{eq:monstergamma}) for the MRI growth rate.}
    \label{tab:sims}
\end{table}

$\eta_{\rm A}$ depends on the ion-neutral collision rate,
\begin{equation}
\gamma_{\rm in} = \frac{\langle\sigma \nu\rangle_{\rm in}}{2m_{\rm i}},
\end{equation}
where $m_{\rm i}$ is the ion mass and $ \langle\sigma \nu\rangle_{\rm in}$ is the rate of ion-neutral interactions. At $T_{\rm i} = T_{\rm n} = 500$~K ion-neutral collisions are dominated by both polarization scattering and charge exchange \citep{lieberman_2005}. We calculate the cross section for resonant charge exchange for argon using Equation (14) in \cite{rapp_1962} and get $\sigma_{\rm cx}=\num{1.4e-18}$~m$^2$. We calculate the cross section for polarization scattering as,
\begin{equation}
    \label{eq:sig_pol}
    \sigma_{\rm p} = 2 \pi Z e \left(\frac{\alpha_{\rm N}}{m_{\rm r}}\right)^{1/2} \frac{1}{v_{\rm i}},
\end{equation}
where $Ze$ is the charge of the ion, $m_{\rm r}$ is the reduced mass, $\alpha_{\rm N}=11.08/a^3$ is the polarizability of argon, $a$ is the Bohr radius, and $v_{\rm i} = \sqrt{2k_{\rm B}T_{\rm i}/m_{\rm i}}$ is the ion velocity \citep{lieberman_2005}. Note that the cross section is inversely proportional to $v$, and therefore $\langle \sigma v \rangle_{\rm p}=\num{6.96e-16}$~m$^3$~s$^{-1}$ will be independent of velocity and therefore temperature. To determine the ion-neutral collision rate we take the sum of the polarization scattering rate and the rate of charge exchange.

The radial profiles for $p$, $n_{\rm e}$, $\chi_{\rm i}$, $T_{\rm e}$, and $u_{\rm r}$ depend on $\eta_{\rm o}$ and $\eta_{\rm a}$, which in turn depend on $p$, $n_{\rm e}$, $\chi_{\rm i}$, and $T_{\rm e}$. Therefore we use an iterative approach. First, we calculate the radial profiles using an initial guess for the diffusivities. Next, we use the average values from the radial profiles to calculate $\eta_{\rm o}$ and $\eta_{\rm a}$, and then we recalculate the radial profiles using the updated diffusivities. We continue to iterate until the change in $\eta_{\rm o}+\eta_{\rm a}$ is less than 20~m$^2$~s$^{-1}$, which should be sufficient because even at the lowest pressures and magnetic field strengths $\eta_{\rm o}+\eta_{\rm a}>400$~m$^2$~s$^{-1}$.

Finally, we define $k_{\rm r} \equiv \pi/(2(r_{\rm 2}-r_{\rm 1}))$ and $k_{\rm z} \equiv \pi/2H$, where $H$ is the height of the apparatus. The instability will appear first at these longer wavelengths, since flows characterized by these wavelengths are less affected by the viscosity which tends to stabilize the flow. We use an averaged orbital frequency, $\Omega \equiv \sqrt{\Omega(r_{\rm 1}) \Omega(r_{\rm 2})}$.

\subsection{Predictions for MRI}
\label{sec:results}
This section presents the numerical solutions to Equation (\ref{eq:monstergamma}) using the methods described in Section \ref{sec:numerical} for the range of parameters outlined in Table \ref{tab:sims}. Over a range of magnetic field strengths and pressures, we use a fiducial power, $P=200$~kW, and aspect ratio, $H/(r_2 -r_1)=2$ ($H=2$~m, $r_1=0.4$~m, $r_2=1.4$~m). We use this fiducial aspect ratio because if $H/(r_2 -r_1)<2$ the MRI cells will be squashed in the axial direction making it difficult to produce. The aspect ratio in an actual protoplanetary disc is $\ll 1$. However, our dispersion relation is a local approximation, so our experiment should be thought of as a localized part of a disc that overall has an aspect ratio $\ll 1$. We use $P=200$~kW because the power needs to be high to reach sufficient ionization fractions (see Appendix \ref{app:nete}). The magnetic fields probed here are several orders of magnitude larger than what is predicted for a protoplanetary disc \citep{Lesur+2022}. However, we are limited on the lower end of our parameter range by the strength of Earth's magnetic field unless a magnetic shield is used.

Figure \ref{fig:diffusive} shows the magnitudes of the diffusivities as a function of pressure at $r_2$ with an initial magnetic field strength of $B_{\rm z}=10^{-3}$~T and $B_{\rm z}=0.01$~T in the left and right panels, respectively. At both magnetic field strengths ambipolar diffusivity is only significant at low pressures, the Hall term is dominant at low to intermediate pressures, and Ohmic diffusivity dominates at higher pressures. Ohmic diffusivity is independent of magnetic field strength, while $\eta_{\rm A} \propto B^2$ and $\eta_{\rm H} \propto  B$. As a result, for $B_{\rm z}=10^{-3}$~T ambipolar diffusion and the Hall effect are weaker and Ohmic diffusivity becomes dominant at roughly $10^2$~mTorr versus 10$^4$~mTorr for $B_{\rm z}=0.01$~T. Therefore, for lower magnetic field strengths, the Hall effect is only dominant at the lowest pressures ($\lesssim50$~mTorr).

For $B_{\rm z}=0.01$~T, ambipolar diffusion can be significant at the lowest pressures, leading to the Hall effect being most dominant at intermediate pressures around 100~mTorr. The Hall effect in this case remains dominant up to around 5~Torr, at which point Ohmic diffusivity becomes dominant. Overall, as magnetic field strength increases, the region of Hall dominance shifts to higher pressures.

Which non-ideal effect is dominant at different pressures and magnetic field strengths dictates at which pressures and magnetic field strengths there is MRI growth, as Figure \ref{fig:g_ideal} shows. The black dots in Figure \ref{fig:g_ideal} show all the values of $-B_{\rm z}$ and $p(r_2)$ used in our calculation regardless of whether MRI growth was found. Although we input an evenly logarithmically spaced grid of $p(r_1)$ from $10^{-3}$ -- $10^4$~mTorr, when we solve Equation (\ref{K}) for $p(r)$, because $u_{\rm r}$ decreases by over 2 orders of magnitude from $r_1$ to $r_2$, the pressure increases with radius (see Figure \ref{fig:profs}). As a result, the minimum $p(r_2) \sim 1$~mTorr. The results we show in this figure are all calculated for negative values of $B_{\rm z}$, or in other words a magnetic field that is anti-aligned with the axis of rotation.

The blue dots in Figure \ref{fig:g_ideal} show values of $-B_{\rm z}$ and $p(r_2)$ at which we get MRI growth with the Hall effect. The orange dots shows values at which we get growth without the Hall effect, that is when we remove the Hall effect from our dispersion relation by setting $\eta_{\rm H}=0$. While doing so does not represent a physical case, it is informative to compare our results with and without the Hall effect to study the impact the Hall effect has on the stability of our proposed experiment.

At low magnetic field strengths, there is only 
MRI growth when $p(r_2)$ is less than a few hundred mTorr, because Ohmic diffusion quickly dominates over the other non-ideal effects. Without the Hall term, there is only MRI growth at these low to intermediate magnetic field strengths, because when the magnetic field is weak, ambipolar diffusion is weak enough to allow MRI growth. MRI growth extends to even lower pressures when the Hall term is included suggesting Hall MRI growth can destabilize a flow otherwise stabilized due to ambipolar diffusivity.

At $-B_{\rm z} \gtrsim 10^{-2}$~T ambipolar diffusion will be very strong, especially at lower pressures. As a result, there will be no MRI growth without the Hall effect. However, the Hall term is also larger for stronger magnetic fields, and so it still becomes dominant over ambipolar diffusion at intermediate pressures and acts to destabilize a flow stabilized due to ambipolar diffusion. Furthermore, because Ohmic resistivity is independent of magnetic field strength, the Hall effect remains dominant up to higher pressures. For $-B_{\rm z} \gtrsim 0.1$~T, MRI growth becomes possible at pressures up to $10-100$~Torr.

We show the logarithm of the number of e-folding times within the residual time of the gas in the experiment in Figure \ref{fig:grate}. We show calculations with the Hall effect in the left panel and without the Hall effect in the right panel. The number of e-folding times ranges from around 4 -- 200 for calculations with the Hall effect and from around 6 -- 40 for calculations without. When the vertical magnetic field is anti-aligned with the axis of rotation, the Hall effect not only increases the parameter-space for MRI instability, but also increases the growth rate of the MRI leading to higher numbers of e-folding times overall.

Figure \ref{fig:visc} shows the logarithm of the Reynolds number (left panel), the magnetic Reynolds number, $R_{\rm m} \equiv \Omega/k^2 \eta_{\rm tot}$ (center panel), and the hall parameter $h \equiv |\eta_{\rm H}|/  \eta_{\rm tot}$ (right panel), as a function of pressure and magnetic field strength for parameters that lead to MRI growth for calculations with the Hall effect. Figure \ref{fig:visc_hist} shows the distributions of $R_{\rm e}$, $R_{\rm m}$, and $h$ for parameters that have MRI growth for calculations with the Hall effect (in blue) and calculations without the Hall effect (in orange). The dashed black lines show the distributions of $R_{\rm e}$, $R_{\rm m}$, and $h$ for parameters that have no MRI growth. The Reynolds number for our range of parameters goes from roughly $R_{\rm e}=1-10^4$ and increases as a function of pressure. MRI growth is most common for intermediate Reynolds numbers. 

The magnetic Reynolds number is much smaller for our range of parameters, ranging from around $R_{\rm m}=10^{-3}-1$. The magnetic Reynolds number decreases as the magnetic field strength increases, because $\eta_{\rm A} \propto B^2$. Without the Hall effect, MRI growth is only possible for the largest values of $R_{\rm m}$ in our calculations. MRI growth is possible for a wider range of $R_{\rm m}$ if the Hall effect is included. 

As Figure \ref{fig:diffusive} suggests, the Hall parameter is strongest at intermediate pressures, where ambipolar diffusion is no longer as strong. At these intermediate pressures the Hall parameter increases as magnetic field strength increases, because $\eta_{\rm H} \propto B$. Because the Hall effect needs to be strong enough to overcome Ohmic resistivity and ambipolar diffusion, MRI growth only occurs when $h>1$.

Figure \ref{fig:g_powers} is the same as Figure \ref{fig:g_ideal}, but for calculations with $P=$10~kW (top panel), and $P=$100~kW (bottom panel). Decreasing the power reduces the range of parameters where MRI growth is possible. Only at $P \gtrsim 100$~kW is it possible to get MRI growth without the Hall effect. At $P=10$~kW it is only possible to get MRI growth with strong magnetic fields. Our calculation with $P=$1~kW had no MRI growth for $H=2$~m, $r_2-r_1=1$~m. However, for a smaller apparatus MRI growth does occur at the largest magnetic field strengths because the power per area increases, which increases the ionization rate.

The results described above are all for cases in which the vertical magnetic field is anti-aligned with the axis of rotation ($B_{\rm z} < 0$). We also performed calculations for which the magnetic field is aligned with the axis of rotation ($B_{\rm z} > 0$). For cases when $B_{\rm z} > 0$, if we do not include the Hall effect in our dispersion relation the results are identical to when $B_{\rm z} < 0$. This symmetry is expected because ambipolar diffusion and Ohmic resistivity are diffusive effects which depend on even powers of $B_{\rm z}$, and hence are independent of sign. On the other hand, when we do include the Hall effect it acts to suppress MRI growth, leading to no MRI growth for our range of parameters. Therefore, to achieve MRI growth it is important to have $B_{\rm z} < 0$.

\section{Prototype experimental results}
\label{sec:prototype}

\begin{figure}
\begin{center}
\includegraphics[width=90mm]{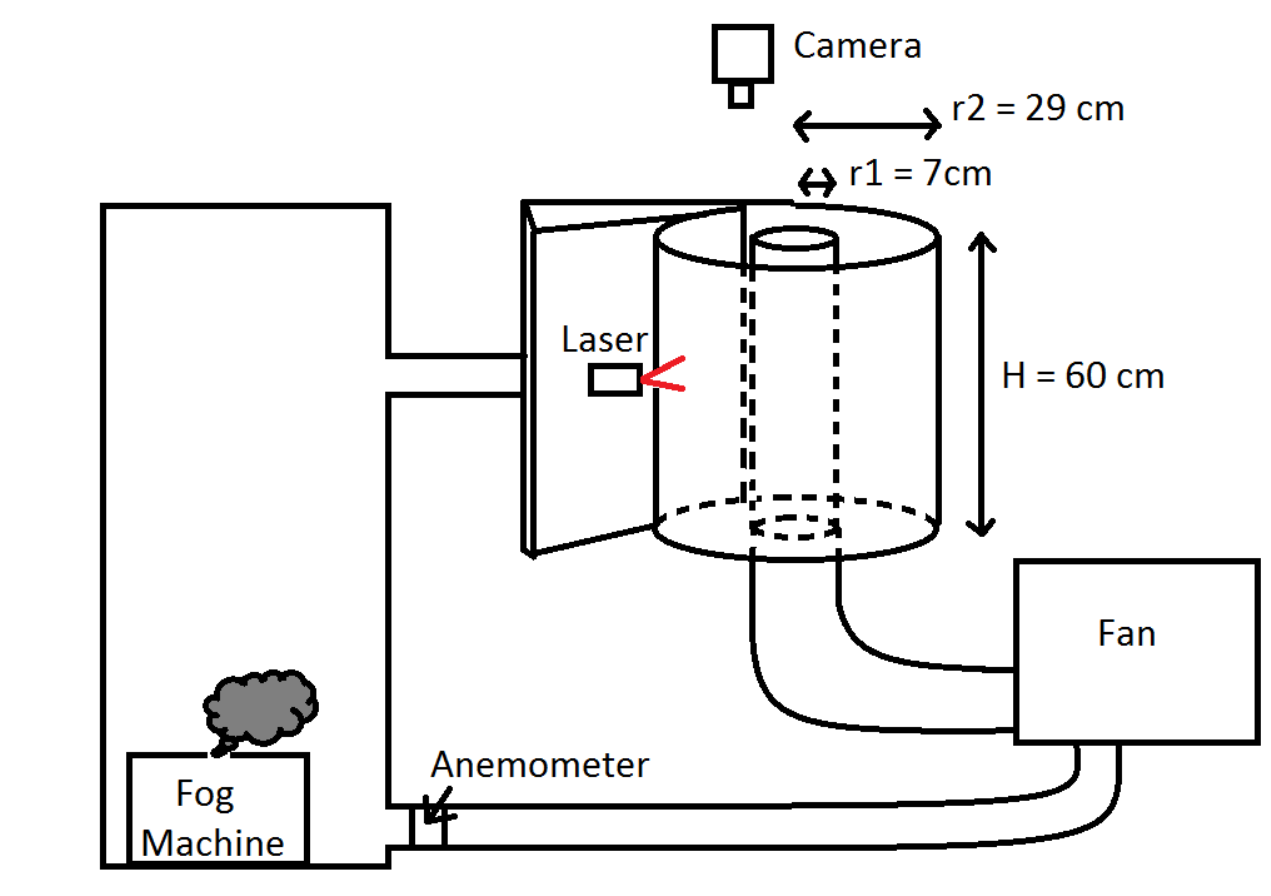}
\caption{A diagram showing the overall setup of our prototype. The air expelled from the fog machine goes through the top tube and enters the gap between the concentric cylinders through a small opening along the edge of the outer cylinder. The fog then rotates in between the cylinders and eventually leaves through holes in the inner cylinder, pulled in by the fan. The exit tube has an anemometer to measure the flux of the air.}
\label{diagram1}
\end{center}
\end{figure}

\begin{figure}
\begin{center}
\includegraphics[width=80mm]{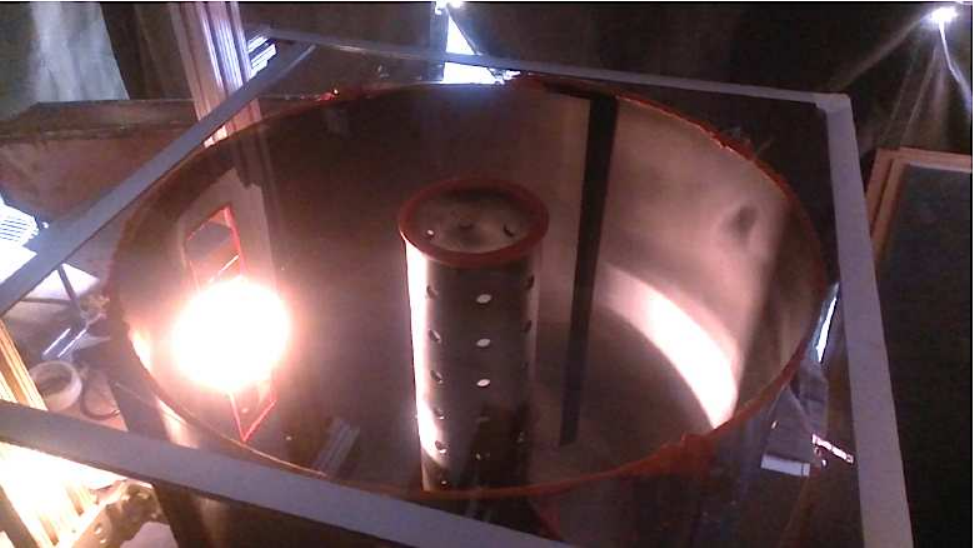}
\caption{Photograph of the apparatus from the top. The inner cylinder with holes is visible, as well as the opening through which the light shines. During operation, we cover all but a small section of the side panel so that light only enters at a specific height.}
\label{photo}
\end{center}
\end{figure}

\subsection{Description of the apparatus}
\setcounter{equation}{0}
To test our experimental design we built a small non-magnetized prototype. The prototype is composed of two concentric cylinders and uses air instead of argon. The outer cylinder of our prototype has a radius of 29 cm and an opening to let air emitted from a fog machine (DYNO-FOG II by American DJ) enter. On another part of the outer cylinder is a small window for the external light, which is provided by a laser sheet. We initially used a construction lamp for the lighting, but found that the light did not penetrate close enough to the inner cylinder. The inner cylinder has a radius of 7.5 cm and it has 80 identical holes (10 ranks of 8 holes) to allow the gas to flow out of the apparatus. The height of the prototype is 60 cm and the upper boundary is plexiglass to allow filming of the flow with a camera. The entire apparatus is airtight.

A gradient of pressure is imposed by a fan placed at the exit pipe. An anemometer is also placed at the exit pipe to measure the flux of the gas. The apparatus scheme is shown in Figure \ref{diagram1}, and Figure \ref{photo} shows a photograph of the prototype. 

To run the experiment, we turn on the laser sheet, start the camera, start the fog machine, and then turn the fan on. As the fog is rapidly sucked into the apparatus, the camera records the fog that is illuminated by the laser sheet. We describe our method of processing this recording in the following section.

\setcounter{equation}{0}
\subsection{Description of the Analysis Methods}
\label{sec:analysis_methods}
To record our experiment, we use a phantom ir300 camera, which is able to record 900 images per second. We optimized the height of the camera ($\sim$22.5cm above the apparatus), the camera lens, and the exposure time (700 $\umu$s). To initially process the video, we use a program specifically for this camera brand: Phantom Camera Control Software. Each frame of the video is processed as a separate image. 

The initial images produced are too low contrast to be analyzed. We improve the contrast in two main steps. First, we compute a local averaged image using 10 images (no more so that we have approximately the same amount of fog in every image). We then divide each image by this local averaged image to normalize the images, improving the contrast. Unfortunately the noise is also increased by this process. Therefore, our second step is to reduce the noise using a local filter. We create this filter by replacing each pixel by a median pixel of itself and its eight neighbor pixels. This process significantly reduces the noise, as the noise is often localized around one pixel. Additionally, it does not significantly affect the actual signal, which is typically spread out over significantly more than eight pixels.

To track the motion of the fog we compute the correlations between different regions of each image for 11 evenly spaced bins in radius and 100 evenly spaced bins in azimuth. For example, if we have areas A and B, the image in which area B most closely resembles area A is selected and then the time difference between the two images and the angular displacement is used to calculate an azimuthal velocity. To calculate the azimuthal velocity we assume the radial velocity can be neglected. This assumption should be valid because for this prototype we expect the $u_{\rm \theta} \gg |u_{\rm r}|$. However, over longer timescales we could also track the radial velocity by looking at the radial displacement instead. We also assume $u_{\rm z}$ is negligible, as the bulk $u_{\rm z}$ should be near zero in our apparatus. However, there could be velocity fluctuations due to turbulence, in particular in the Ekman layers. Here, we avoid the Ekman layers where turbulence should be largest and only measure $u_{\rm \theta}$ at intermediate heights where it should be much larger than the other velocity components. In our apparatus the camera is located at a specific height and only records movement in the azimuthal direction. Future experiments could rotate the camera to determine $u_{\rm z}$, which could be used to verify that $u_{\rm z}$ is negligible.

\subsection{Experimental Results}

\begin{figure}
\begin{center}
\includegraphics[width=1.\linewidth]{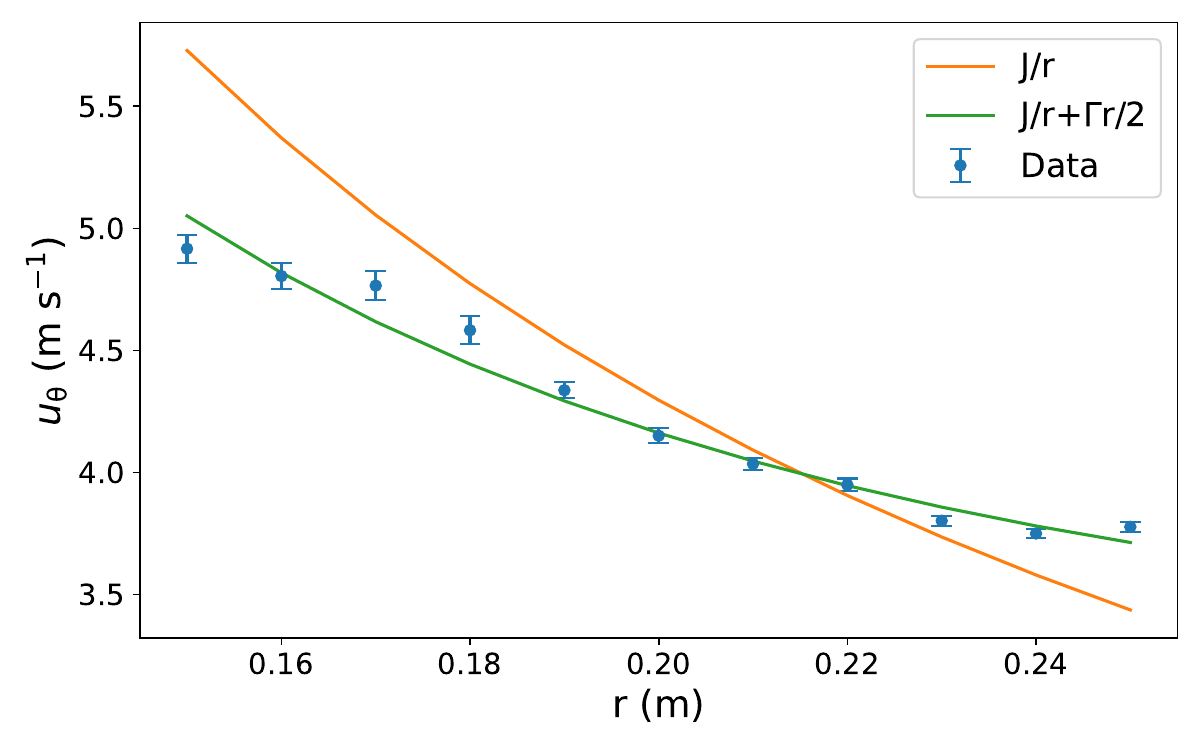}
\caption{The azimuthal velocity as a function of radius measured at a height of 38~cm in our prototype experiment. We fit a model $u_{\rm \theta}=J/r$ to our data, shown in orange, and find $J=0.86 \pm 0.02$~m$^2$~s$^{-1}$ with reduced-$\chi^2 = 70$. We also fit a model $u_{\rm \theta}=J/r+\Gamma r/2$ to our data, shown in green, and find $J=0.66 \pm 0.02$~m$^2$~s$^{-1}$ and $\Gamma=8.54 \pm 0.66$~s$^{-1}$ with reduced-$\chi^2 = 3.6$.}
\label{fig:reg}
\end{center}
\end{figure}

Figure \ref{fig:reg} shows the azimuthal velocity as a function of radius in our experimental device. We show the mean of the distribution of azimuthal velocities for each radius bin in blue. The errors are calculated as the standard deviation of the azimuthal velocities in each radial bin divided by the number of measurements in that bin. The errors are largest at radii closer to the inner cylinder, because the inner radii are farthest from the light (see Figure \ref{diagram1}). Insufficient illumination appears to be the largest source of error in our data. The flux measured by the anemometer in the exit pipe is $D = 0.11$ m$^{3}$ s$^{-1}$. Using Equation (\ref{Kmu}) and the kinematic viscosity of air at room temperature and normal pressure, $\nu \approx 1.5 \times 10^{-5}$ m$^{2}$ s$^{-1}$, we find $K/\mu = \num{1.95e3}$. Therefore, $K \gg \mu$ and we can use Equation (\ref{uthetaidk}) to find a fit to the azimuthal velocity.

 We first make the assumption that $F_{\rm \theta}=0$, which means equation (\ref{uthetaidk}) becomes $u_{\rm \theta}=J/r$. The orange line in Figure \ref{fig:reg} shows this function for $u_{\rm \theta}$ fit to our data. We find $J=0.86\pm 0.02$~m$^2$~s$^{-1}$ with a large reduced-$\chi^2 = 70$. If we assume instead $F_{\rm \theta}<0$, equation (\ref{uthetaidk}) becomes $u_{\rm \theta}=J/r + \Gamma r/2$ . The green line in Figure \ref{fig:reg} shows this function fit to our data. We find $J=0.66 \pm 0.02$~m$^2$~s$^{-1}$ and $\Gamma=8.54 \pm 0.66$~s$^{-1}$. With these values reduced-$\chi^2 = 3.5$, which is a factor of 20 smaller than for our fit with $F_{\rm \theta}=0$, suggesting $F_{\rm \theta}<0$ in our experiment. 

\begin{figure}
\begin{center} 
\includegraphics[width=1.\linewidth]{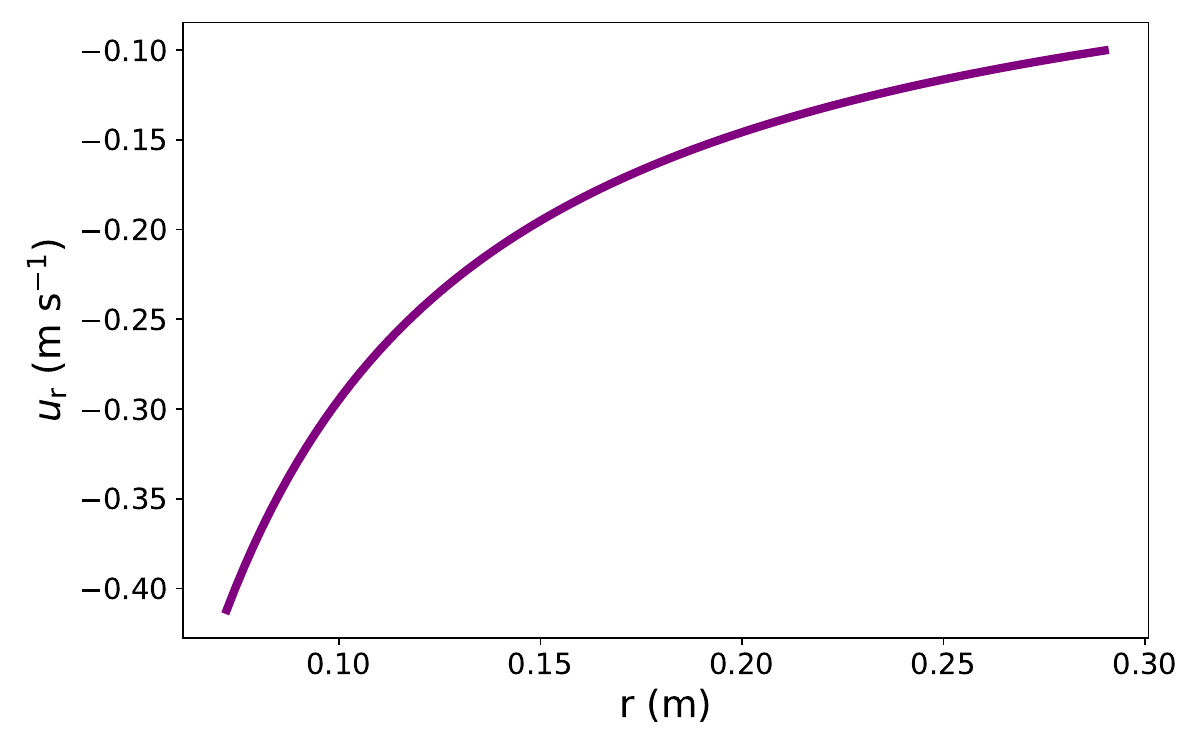}
\caption{The radial velocity as a function of radius in our prototype. To find $u_{\rm r}(r)$ we integrate equation (\ref{ur2}) using the $J$ and $\Gamma$ from our fit in Figure \ref{fig:reg}.}
\label{fig:VrComparison}
\end{center}
\end{figure}

We show the radial velocity as a function of radius in Figure \ref{fig:VrComparison}. We derive this radial velocity profile using our fitted values of $J$ and $\Gamma$ and the flux measured by the anemometer, $D$, to integrate equation (\ref{ur2}). We find that in our prototype the radial velocity is at least an order of magnitude smaller than the azimuthal velocity, which validates our assumption that the radial velocity can be neglected in order to determine the azimuthal velocity, as described in Section \ref{sec:analysis_methods}.

\begin{figure}
\begin{center} 
\includegraphics[width=1.\linewidth]{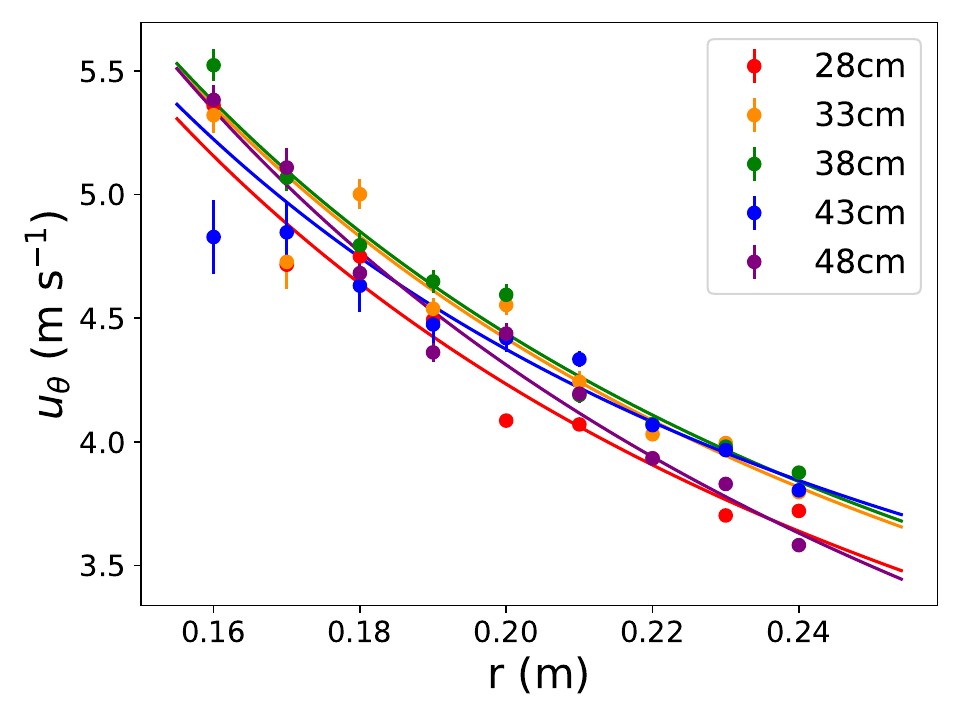}
\caption{Radial profiles of the azimuthal velocity taken at multiple heights. The profiles for the top and bottom height differ the most from the other profiles at the largest radii, but are still consistent to within $\sim 20\%$.}
\label{fig:heightcomp}
\end{center}
\end{figure}

Figure \ref{fig:heightcomp} again shows the azimuthal velocity as a function of radius, but now for several different heights inside the apparatus. The radial profiles for the various heights are mostly consistent with each other. The profiles for heights of 28~cm and 48~cm differ the most from the other profiles at the largest radii. However, they are still consistent to within around 20\%.

We use Equation (\ref{Reynolds}) to get a Reynolds number for our system, $Re = 4.36 \times 10^{5}$. We also use Equation (\ref{alpha2}) to find $\alpha =  2.85\times 10^{-6}$ in our prototype, which is many orders of magnitude smaller than the $\alpha$ estimated from the accretion rates in protoplanetary discs, $\alpha_{\rm ppd} \approx 10^{-2}$ \citep{hartmann98}. Converting the kinematic viscosity of air at room temperature and normal pressure to $\alpha$ we get, $\alpha_{\rm air}=\num{4.2e-9}$. Therefore for our prototype we have $\alpha_{\rm air} \ll \alpha \ll \alpha_{\rm ppd}$, which means the viscosity is small, but non-negligible.

This small viscosity could be due to either hydrodynamic turbulence or Ekman effects. The residual time of the gas in the device, $\sim 1.4$~s, is around an order of magnitude shorter than the Ekman time \citep{kageyama04}, 
\begin{equation}
    \label{eq:ekman}
    \tau_{\rm E} \approx \frac{H}{2 \sqrt{\nu \Omega}} \approx 12\rm{~s}.
\end{equation}
Therefore, hydrodynamic turbulence must play some role in driving the viscosity. From Figure \ref{fig:reg} we can see that at inner radii the azimuthal velocity is lower than it would be if angular momentum is conserved (i.e. $F_{\rm \theta}=0$), and at outer radii the azimuthal velocity is somewhat higher, suggesting that this low level of turbulence is leading to angular momentum being transferred outwards, which in turn leads to accretion.

\section{Discussion and Conclusions}
\label{sec:discuss}
In this paper we have examined the feasibility of obtaining non-ideal MHD MRI growth in a swirling argon gas laboratory experiment meant to resemble a protoplanetary disc. First, we derive the hydrodynamic equilibrium flow of our proposed experiment. Next, we derive the dispersion relation for MRI growth with Ohmic resistivity, ambipolar diffusion, and the Hall effect. We then solve this dispersion relation numerically for a simulated experiment over a range of parameters. We find that for $P=200$~kW MRI growth is possible for $10$~mTorr $\lesssim p(r_2) \lesssim 1$~Torr and $10^{-3}$~T $\lesssim B_{\rm z} \lesssim \num{2e-2}$~T without the Hall effect and for $1$~mTorr $\lesssim p(r_2) \lesssim 100$~Torr and $10^{-3}$~T $\lesssim |B_{\rm z}|\lesssim 0.2$~T with the Hall effect, when the magnetic field is anti-aligned with the axis of rotation. In cases where there is MRI growth several e-folding times, sometimes over 100, will occur within the residual timescale.

At lower powers it is difficult to get MRI without the Hall effect and the range of parameters over which MRI growth is possible with the Hall effect, decreases as power decreases. In general the range of parameters over which MRI growth is possible is larger when we include the Hall effect in our dispersion relation than when we do not.

We also describe our un-magnetized prototype. Using only an anemometer at the exit pipe and the method described in Section \ref{sec:analysis_methods} to fit a model to $u_{\rm \theta}(r)$, we are able to determine the radial velocity, Reynolds number, and $\alpha$-parameter of our experiment. We are also able to validate our assumptions that $K \gg \mu$ and that our flow is constant as a function of height to within 20\%. 

We find that in our prototype the azimuthal velocity depends on a frictional force $F_{\rm \theta}<0$. While the corresponding $\alpha$-parameter is much lower than that of a protoplanetary disc, it is still significant compared to the viscosity of air at room temperature and normal pressure, and appears to lead to outward angular momentum transport. 

We can use this $\alpha$-parameter as a baseline to compare to a future magnetized experiment, like the one proposed here. However, because it is a prototype, the apparatus described in Section \ref{sec:prototype} has a small radial extent and low azimuthal velocities compared to our proposed magnetized experiment. In particular, the azimuthal velocity in our proposed experiment will be supersonic at standard atmospheric pressure, while it is subsonic in our prototype. As a result, there will be several differences between the two. For example, the Reynolds number in our proposed experiment will be smaller than in our prototype (see Figure \ref{fig:visc_hist}), because even though the azimuthal velocity in our proposed experiment is over 100 times larger, the density will be several orders of magnitude smaller. Another example is that, as Figure \ref{fig:profs} shows, the pressure in our proposed experiment will increase by around 2 orders of magnitude from the inner to the outer radius, as opposed to being nearly constant in our prototype. Because the gas temperature is assumed to be constant in our proposed experiment, this increase in pressure suggests the steady-state flow may not be incompressible as we have assumed when deriving the dispersion relation in Section \ref{sec:disp}. Future work should derive a dispersion relation for compressible gas.

Our calculations show that the Hall effect enhances the range of parameters for which we get MRI growth and the rate of MRI growth when the magnetic field is anti-aligned with the rotation axis ($B_{\rm z} < 0$). Our results are consistent with \cite{wardle99} and \cite{bai15} who showed the Hall effect can enhance the MRI when $B_{\rm z} < 0$. 

On the other hand, we found that all MRI growth is suppressed when the magnetic field and rotation axis are aligned. MRI growth is suppressed because the Hall effect tends to stabilize the MRI when $B_{\rm z} > 0$, and also because our proposed experiment is stable to the Hall-Shear Instability \citep[HSI,][]{kunz08}. Our proposed experiment is HSI stable because our shear rate is very low. As a result, HSI would only be achievable at high pressures and low magnetic field strengths, where in our case Ohmic resistivity is strong enough to suppress the HSI.

While our proposed experiment will probe whether MRI is possible in a non-ideal MHD gas disc with similar gas number densities and gas pressures to those assumed for a protoplanetary disc, we are still limited in our ability to fully replicate the environment of a protoplanetary disc. First, we are using argon gas for our experiment, when a protoplanetary disc is primarily composed of hydrogen and helium gas, as well as many species of dust. Second, our magnetic field strength will be several orders of magnitude stronger than the predicted magnetic field strengths for protoplanetary discs, because we are limited by the Earth's magnetic field, $B \sim 10^{-4}$~T. Also, while we probe values of plasma $\beta = P_{\rm th}/P_{\rm B} \gtrsim 10^4$ as are expected for a protoplanetary disc \citep{Lesur+2022}, our calculations only produce MRI in cases where $\beta \leq 10$. 

In addition, we have used an aspect ratio, $H/r >1$, when a protoplanetary disc will have $H/r \ll 1$. However, we have used a local approximation to derive our dispersion relation and so our experiment can be thought of as a local approximation for a global disc that may as a whole have $H/r \ll 1$. Finally, the ionization fractions ($\chi_{\rm i} \approx 10^{-5}$ -- $10^{-1}$) in our proposed experiment are many orders of higher than those for a protoplanetary disc \citep[$\chi_{\rm i} \approx10^{-13}$ -- $10^{-6}$,][]{Lesur+2022}. However, these ionization fractions are similar to those in the interstellar medium, which is also subject to non-ideal MHD MRI turbulence. 

While there are key differences between our proposed experiment and a protoplanetary disk, it will still be useful to show whether the MRI can grow in weakly-ionized gas subject to non-ideal effects. In addition, with this experiment we should be able to explore the physics of the MRI in a parameter study complementary in some respects (e.g. Reynolds number) to parameter studies performed with numerical simulations, and without several of the simplifications made in these simulations. Furthermore, the results of this proposed experiment can be used to compare to analytic and numerical solutions and provide insight on the significant physics to include.

The next step should be to design, and if feasible to build, an experiment of radius on the order of $r_2=1.4$~m with a power on the order of $P=200$~kW. Our calculations suggest that setting the pressure at the outer radius of the apparatus to around $p(r_2)=100$~mTorr and generating a vertical magnetic field with a magnitude of around $B_{\rm z}=0.01$~T that is anti-aligned with the axis of rotation is ideal for trying to produce MRI in the presence of non-ideal MHD effects. Doing so would enable us to directly observe the MRI under conditions similar to those of protoplanetary discs, which could provide insight to the origin of angular momentum transport in these discs.

\section*{Acknowledgements}
We thank the anonymous referee for their helpful comments which improved the clarity of this paper. A.S. is supported by a fellowship from the NSF Graduate Research Fellowship Program under Grant No. DGE-1656466. H.J. acknowledges support by the Laboratory Directed Research and Development (LDRD) Program at Princeton Plasma Physics Laboratory under the U.S. Department of Energy Contract No. DE-AC02-09CH11466. J.G. acknowledges support from NASA grant 17-ATP17-0094 and NSF grant AST-2108871. H.J. thanks Jill Foley, Michael Kennelly, Mark Nornberg, Stefan Gerhardt, Brandon Fetroe, Yevgeny Raitses, Jenna Kefeli, Hans Rinderknecht, Enrique Merino, Igor Kaganovich, Joey McDonald, Alex Gurak, Erdem Oz, Cami Collins, Courtney Kaita, Bob Cutler, Eric Edlund, Austin Wang, Kyle Kremer, Owen Williams, Lex Smits, Tapash Sarkar, Matthew Basile, Samuel Greess, Erik Gilson, and Daniel Gift for their contributions over many years to this project.

\section*{Data Availability}
The data underlying this article will be shared on reasonable request to the corresponding author.

\bibliographystyle{mnras}
\bibliography{MRI.bib}

\appendix

\section{Solution for the Azimuthal Velocity}
\label{app:sol}
To find the azimuthal velocity we first note that since $\Gamma$ has no $r$ dependence, we can write the homogeneous form of equation (\ref{theta2}) as,
\begin{equation} \label{appendixB1}
\frac{\partial{u_{\rm \theta}}}{\partial{r}} + \frac{u_\theta}{r} + \frac{\mu}{K}\left[\frac{\partial}{\partial{r}}\left(r\frac{\partial{u_{\rm \theta}}}{\partial{r}}\right) - \frac{u_{\rm \theta}}{r}\right]=0,
\end{equation}
where we have set $\Gamma=0$. This equation is an equidimensional linear (or Cauchy-Euler) ordinary differential equation (ODE), which have power-law solutions. If we define $f$ such that $u_{\rm \theta} = f/r$, equation (\ref{appendixB1}) can be written as,
\begin{equation}
f'' = \frac{f'}{r}\left(1-\frac{K}{\mu}\right),
\end{equation}
which we can integrate to get,
\begin{equation}
f' = a r^{1-\frac{K}{\mu}},
\end{equation}
and integrate again to get,
\begin{equation}
u_{\rm \theta} = \frac{J}{r} + \frac{a}{2-\frac{K}{\mu}}r^{1-\frac{K}{\mu}}.
\end{equation}
Thus we have a general solution to equation (\ref{appendixB1}), if $K/\mu \neq 2$. For  $K/\mu \approx 2$, $J/r$ is the only linearly independent power-law solution. In this case we find a second linearly independent solution in the form of a logarithm times the first solution, which gives us,
\begin{equation}
u_{\rm \theta} = \frac{J}{r} + \frac{a}{r}\ln(r/r_1).
\end{equation}

We now look for a particular solution to 
\begin{equation}
\frac{\partial{u_{\rm \theta}}}{\partial{r}} + \frac{u_{\rm \theta}}{r} + \frac{\mu}{K}\left[\frac{\partial}{\partial{r}}\left(r\frac{\partial{u_{\rm \theta}}}{\partial{r}}\right) - \frac{u_{\rm \theta}}{r}\right] = \Gamma.
\end{equation}
Solutions to Cauchy-Euler inhomogenous ODEs have particular solutions $\propto r$.
We take $r \Gamma / 2$ as a particular solution to this equation. Therefore our solution to equation (\ref{theta2})  is,
\begin{equation}
u_{\rm \theta} = \frac{J}{r} + \frac{r \Gamma}{2} + \frac{a}{2-\frac{K}{\mu}}r^{1-\frac{K}{\mu}},
\end{equation}
where the last term would be replaced by $ar^{-1}\ln(r/r_1)$ if $K/\mu \approx 2$.

We now look at the two limiting cases, $K \ll \mu$ and $K \gg \mu$. The first case, $K \ll \mu$, is straightforward. $K/\mu \rightarrow 0$ and we get
\begin{equation}
    \label{eq:kllmu}
    u_{\rm \theta} = \frac{J}{r} + \frac{r\Gamma}{2},
\end{equation}
where $\Gamma \rightarrow \Gamma + a$. 

For $K \gg \mu$, we define the function,
\begin{equation}
    \label{eq:kggmu}
    f(x,y) = \frac{x^{(1-y)}}{2-y},
\end{equation}
where $y\equiv K/\mu$ and $x\equiv ra^{(1/(1-K/\mu))}$, which are both dimensionless. The question is, what is the limit of $f(x,y)$ when $y\rightarrow\infty$ at fixed $x$. If $|x|>1$, $f(x,y)\rightarrow 0$, when $y\rightarrow\infty$. If $|x|<1$, $f(x,y)\rightarrow -\infty$, when $y\rightarrow\infty$, which would mean a supersonic flow. Therefore, it is necessary for $a=0$ to avoid a supersonic flow. This solution then gives us Equation (\ref{eq:kllmu}) for both limiting cases.


\section{A model to illustrate the physical meaning of $F_\theta$}
\label{app:ftheta}

To illustrate the physical meaning of $F_{\rm \theta}$, we relax our assumption that $u_{\rm r}$ and $u_{\rm \theta}$ are independent of $z$. The azimuthal component of the Navier-Stokes equation is now,
\begin{equation}
\label{eq:ansz}
\rho\left[u_{\rm r}\frac{\partial u_{\rm \theta}}{\partial r} + \frac{u_{\rm r}u_{\rm \theta}}{r}\right] =
 \mu \left[\frac{1}{r}\frac{\partial}{\partial r}\left(r\frac{\partial u_{\rm \theta}}{\partial r}\right) - \frac{u_{\rm \theta}}{r^{2}}
 +\frac{\partial^{2}u_{\rm \theta}}{\partial z^{2}}\right]
 \end{equation}
We can then write the solution to this equation as,
 \begin{equation}
 \label{eq:uthetaz}
 u_{\rm \theta} = \left(\frac{J}{r} + f(r)\right)g(z)
 \end{equation}
 
Substituting equation (\ref{eq:uthetaz}) into equation (\ref{eq:ansz}), we can separate the two variables and get,
 \begin{equation}
 f' + \frac{f}{r} + \frac{\mu}{K}\left( f' + rf'' - \frac{f}{r}\right) = -\frac{\mu J g''}{K g} = \Gamma,
 \end{equation}
 where we have assumed $f(r) \ll J/r$. $\Gamma$ is still a constant because the left part of the equation is only $r$-dependent and the middle part is only $z$-dependent. The solution for the left hand side is the same as in Appendix \ref{app:sol}. Solving for $g(z)$ yields,
 \begin{equation}
 g(z) = a \cos \left(\sqrt{\frac{-\Gamma  K }{\mu J} }z\right) +  b \sin \left(\sqrt{\frac{-\Gamma K}{\mu J} }z\right).
 \end{equation}
 
 If we take the coordinate origin at the center of the apparatus we have $g(z) = g(-z)$ and $g(\frac{H}{2})=0$. These boundary conditions give us $b=0$ and $\Gamma =\frac{\mu J \pi^{2}}{K H^{2}}$. 
 Thus we get,
 \begin{equation} \label{longutheta}
 u_{\rm \theta}(r,z) = \left(\frac{J}{r} +\frac{J \pi^{3} \nu r}{D H}+ \frac{a}{2-\frac{D}{2\pi H \nu}} r^{1-\frac{D}{2\pi H \nu}}\right)\cos\left(\frac{\pi z}{H}\right)
 \end{equation}
after using $K = \frac{D \rho}{2 \pi H}$. A numerical estimation with our parameters confirms that the assumption $f(r) \ll \frac{J}{r}$ is valid. To simplify we take the limiting case as in Equation (\ref{eq:kllmu}),
\begin{equation} \label{shortutheta}
u_{\rm \theta}(r,z) = \left(\frac{J}{r} +\frac{J \pi^{3} \nu r}{D H}\right)\cos\left(\frac{\pi z}{H}\right).
\end{equation}

This model is not fully correct because it does not take into account $u_{\rm z}$, which has to exist because of the Ekman circulation. We use it here only to illustrate a few intuitive points about the physical meaning of  $F_{\rm \theta} = -K\Gamma/r$. First, $\Gamma$ is proportional to $\nu$, which is characteristic for a viscous drag force. Second, $\Gamma$ decreases as $H$ increases, meaning it is greatest at the center of the apparatus. Finally if $D$ increases, the Reynolds number increases and thus $\Gamma/J$ decreases. Therefore $F_{\rm \theta}$ represents viscous drag against the vertical boundaries, including Ekman circulation.

\section{Model to compute electron density and temperature}
\label{app:nete}
In this section, we briefly summarize the model derived in \cite{lieberman_2005} for electron temperature and number density. We use this model to compute these values in our numerical calculations. As a simplification, we only take vertical loss into account, neglecting radial loss to the axial magnetic field. In this Appendix all temperatures are given in electronvolts.

\subsection{Computation of $T_{\rm e}$}
For a cylindrical discharge model in the low-to-intermediate ion mean free path regime, we can determine the electron temperature from particle balance by equating the total surface particle loss to the total volume ionization,
\begin{equation}
    \label{eq:equate}
    n_{\rm e} u_{\rm B}A = K_{\rm iz} n_{\rm n}n_{\rm e} \pi r_{21}^2H,
\end{equation}
where $u_{\rm b} = \sqrt{T_{\rm e}/m_{\rm i}}$ is the characteristic velocity at which ions flow out of the apparatus known as the Bohm velocity, $A=2\pi r_{21}^2 H h_{\rm H}$ is the area for particle loss, $K_{\rm iz}$ is the ionization rate constant, and $r_{21}=r_2-r_1$. 

\begin{equation}
h_{\rm H} = \frac{0.86}{\sqrt{3 + \frac{H}{2 \lambda_{\rm i}} + (0.86 H v_{\rm i}/(\pi D_{\rm a}))^2 }},
\end{equation}
is the edge to center density ratio \citep{lee1995}, where $\lambda_{\rm i}=(1/n_{\rm i} \sigma_{\rm i})$ is the ion mean free path, $D_{\rm a} \approx D_{\rm i} T_{\rm e}/T_{\rm i}$ is the macroscopic ambipolar diffusion coefficient, and $\sigma_{\rm i}$ is the ion collision cross-section. The ambipolar diffusion coefficient is, $D_{\rm i} = \pi/8 \lambda_{\rm i}^2 \nu_{\rm i}$, where $\nu _{\rm i}=v_{\rm i}/\lambda_{\rm i}$ is the ion viscosity and $v_{\rm i}=\sqrt{8T_{\rm i}/\pi m_{\rm i}}$. Using these definitions we can simplify the above equation to,
\begin{equation}
    \label{eq:hhfin}
    h_{\rm H} = \frac{0.86}{\sqrt{3 + \frac{H}{2 \lambda_{\rm i}} + 0.19\left(\frac{H}{\lambda_{\rm i}}\right)^2\frac{T_{\rm i}}{T_{\rm e}}}}.
\end{equation}

We can simplify equation \ref{eq:equate} to obtain,
\begin{equation}
    \frac{K_{\rm iz}(T_{\rm e})}{u_{\rm B}(T_{\rm e})}=\frac{1}{n_{\rm n} d_{\rm eff}},
\end{equation}
where $d_{\rm eff} = H/2h_{\rm H}$ is an effective plasma size for particle loss. Using the electron temperature dependence of $K_{\rm iz}$ and $u_{\rm b}$ we get (see Figure 10.1 of \cite{lieberman_2005}),
\begin{equation}
\label{eq:Te}
T_{\rm e} = \frac{1 \rm{~eV}}{0.167 + 0.1315 \log_{\rm 10}(n_{\rm n}d_{\rm eff})},
\end{equation}

Because $K_{\rm iz}$ varies exponentially with $T_{\rm e}$, $T_{\rm e}$ only varies logarithmically with $n_{\rm n}$ and $d_{\rm eff}$. Figure \ref{fig:profs} shows that $T_{\rm e}$ varies by only a factor of 2 from the inner to the outer radius of our proposed experiment, even though the gas density increases by around 2 orders of magnitude. In addition, the temperature depends only on particle conservation, not on the electron density. Thus it is independent of the input power.

To accurately compute $T_{\rm e}$ in our numerical calculation, we first make an initial guess for $T_{\rm e}$ and use this value to compute $h_{\rm H}$. We then use this value for $h_{\rm H}$ to re-compute $T_{\rm e}$. We iterate through this process until $T_{\rm e}$ converges to within $10^{-3}$~eV.

\subsection{Computation of $n_{\rm e}$}
We use the electron temperature to compute the electron number density. We assume Maxwellian electrons absorbing electrical power, $P$, and equate the total power absorbed to the total power lost,
\begin{equation}
    \label{eq:pbal}
    P=e n_{\rm e} u_{\rm B} A E_{\rm tot},
\end{equation}
where $E_{\rm tot} = E_{\rm i}+E_{\rm e}+E_{\rm c}$ is the total energy lost, which depends on the collisional energy lost per electron-ion pair created, $E_{\rm c}$, as well as the kinetic energy carried by electrons and ions to the walls, $E_{\rm e}$ and $E_{\rm i}$, respectively. 

For Maxwellian electrons $E_{\rm e}=2T_{\rm e}$. $E_{\rm i}$ is the sum of the ion energy entering the sheath ($1/(2m_{\rm i})u_{\rm B}^2=T_{\rm e}/2$) and the energy gained by the electron as it traverses the sheath ($\sim 4.7 T_{\rm e}$ for an insulating wall). We approximate $E_{\rm c}$ as,
\begin{equation}
    \log_{\rm 10}E_{\rm c} = 2.18 e^{-3 \log_{\rm 10}T_{\rm e}} + 1.28,
\end{equation}
which accounts for the loss of electron energy due to ionization, excitation, and polarization scattering against neutral atoms \citep{lieberman_2005,Gudmundsson2002}.

We can solve equation (\ref{eq:pbal}) for $n_{\rm e}$ to get,
\begin{equation}
\label{eq:ne}
n_{\rm e} = \frac{P}{eU_{\rm b}A_{\rm eff}E_{\rm tot}}.
\end{equation}

\label{lastpage}
\end{document}